\documentclass[final]{article}

\usepackage{braket,amsfonts}

\usepackage{array}
\usepackage{doi}
\usepackage{CJKutf8}
\usepackage[caption=false]{subfig}
\usepackage{siunitx}
\usepackage{pgfplots}

\usepackage{amsmath}
\usepackage{amssymb}
\usepackage{amsthm}
\usepackage{algorithm}
\usepackage{algorithmic}

\usepackage{graphicx,epstopdf}

\usepackage{amsopn}

\usepackage{xspace}
\usepackage{bold-extra}
\usepackage[most]{tcolorbox}
\usepackage{gensymb}

\colorlet{texcscolor}{blue!50!black}
\colorlet{texemcolor}{red!70!black}
\colorlet{texpreamble}{red!70!black}
\colorlet{codebackground}{black!25!white!25}

\newtheorem{lemma}{Lemma}[section]
\newtheorem{theorem}{Theorem}[section]

\title{Quantitative analysis of cell size control mechanisms
\footnote{The work was supported by NSFC of grant 12331018.}}

\date{\today}

\author{SHUQI FAN\thanks{School of Mathematical Sciences, Center for Applied Mathematics, Tiangong University, Tianjin 300387, China (Email: fanshuqi@tiangong.edu.cn}
\and JINZHI LEI\thanks{School of Mathematical Sciences, Center for Applied Mathematics, Tiangong University, Tianjin 300387 (Email: jzlei@tiangong.edu.cn).}}

\begin{document}
\maketitle

\begin{abstract}
Cell size control is crucial for maintaining cellular function and homeostasis. In this study, we develop a first-order partial differential equation model to examine the effects of three key size control mechanisms: the sizer, timer, and adder. Each mechanism is incorporated into the model through distinct boundary conditions. Exact solutions for these mechanisms are derived using the method of characteristics, allowing us to explore how the steady-state size distribution depends on control parameters. Additionally, individual-cell-based stochastic simulations are performed to validate our theoretical findings and investigate the size distribution under various conditions. This study provides new insights into the quantitative dynamics of cell size regulation, highlighting the underlying mechanisms and laying the groundwork for future theoretical and experimental work on size homeostasis in biological systems. 
\end{abstract}


\vspace{0.5cm}
\textbf{MSCcodes:\ \ }92C37, 35F15, 35C05


\section{Introduction}

Cell size varies widely across species and cell types, and the appropriate size is crucial for the efficient functioning of cells in their respective contexts \cite{gong2022measuring, jones2023characterization,  meizlish2021tissue}. In unicellular organisms, cell size has been shown to correlate with cellular fitness in fluctuating environments \cite{Monds:2014aa}. In multicellular organisms, where environmental conditions are more stable, proper cell size is essential for specialized cellular functions \cite{Kwon:2001aa,Latronico:2004aa}. As such, cells of a given type must maintain a characteristic size to function efficiently within their ecological or organismal context \cite{Amodeo:2016aa,Westfall:2017gd}. Achieving this size homeostasis requires tight coupling between cell growth and division, ensuring that cell size variance remains controlled across successive cell cycles \cite{Amodeo:2016aa,Chadha:2024aa,Rhind:2021aa,Willis:2017aa}. Despite extensive studies on the cell cycle regulatory network, the mechanisms linking cell growth to division--and thereby controlling cell size--remain elusive \cite{ginzberg2015cell,Turner:2012aa}. 

The coupling of cell growth and division is central to cell size control and may occur at different stages of the cell cycle. The cell cycle encompasses a sequential series of events leading to cell division, including cell growth (G1, S, and G2 phases), DNA duplication (S phases), organelle duplication (G2 phase), and partition of cellular components during mitosis (M phase). Cell growth occurs continuously throughout the cycle, while cell size decreases discontinuously after division. Consequently, cell sizes following division may exhibit significant variability. However, robust cell size control mechanisms ensure a relatively stable size distribution for proliferating cells of the same type \cite{ginzberg2015cell,taheri2015cell,Tzur2009Cell}. 

The cell cycle includes several checkpoints to ensure proper progression, with three major ones: the G1 checkpoint (Start or restriction checkpoint), the G2/M checkpoint, and the metaphase-to-anaphase transition (spindle checkpoint). Cell size control often integrates with these checkpoints to coordinate cell growth and division. For instance, in budding yeast, size control predominantly occurs at the G1/S transition \cite{barber2020cell,palumbo2016whi5,talia2007effects}, while in fission yeast, it is primarily regulated at the G2/M transition \cite{wood2013pom1}. 

At the molecular level, cell size control involves multiple intracellular processes. For example, the balance between ribosomal activity and division protein synthesis is critical for coordinating cell growth and division rates \cite{serbanescu2020nutrient}. Gene regulatory networks also play a key role in sensing and regulating cell size \cite{Lloyd:2013aa,otto2010signalling}. In mammalian cells, the mTOR pathway acts as a central signaling hub, integrating nutrient and growth factor signals to regulate protein synthesis and cell size \cite{laplante2012mTOR,locasale2011Metabolic}. Despite these insights, the molecular mechanisms governing cell size control remain one of the fundamental open questions in cell biology.

Previous studies have identified three mechanisms of cell size control: sizer, timer, and adder. The sizer mechanism posits that cells divide upon reaching a critical size, requiring size monitoring \cite{Facchetti:2017aa,koch1962model}. The timer mechanism suggests that cells grow for a fixed duration before division  \cite{donnan1983cell}, while the adder mechanism asserts that cells add a constant volume before divisions \cite{campos2014constant, taheri2015cell}. The sizer and adder mechanisms have been validated in bacteria and yeast through experimental observations \cite{chandler2017adder,soifer2016single, tanouchi2015noity}, while hybrid strategies exhibiting sizer-like or timer-like behaviors have also been reported \cite{Miotto2024A}. These mechanisms, combined with the growth law and cell cycle dynamics, have inspired numerous theoretical models and experimental studies \cite{Cadart:2018aa,DONACHIE1968Relationship,Iyer-Biswas:2014aa,jia2021cell,POWELL1964A,Proulx-Giraldeau:2022aa}.

The issue of cell size control presents a significant mathematical challenge for quantitatively modeling and analyzing the evolution of cell size distributions. Over the years, computational and mathematical models have been developed to describe the effects of various cell size control mechanisms. Early studies introduced partial differential equation (PDE) models to describe the evolution of cell number (density) as a function of cell age and size, often assuming symmetric division \cite{Bell1968Cell, Bell1967Cell, Diekmann1984On, James1971A, m1925applications}. These models laid the foundation for understanding cell growth and division dynamics. 

The growth and fragmentation of cell populations have also been described using modified Fokker-Planck equations, as shown by: 
\begin{equation}
\label{eq:1.1}
\dfrac{\partial\ }{\partial t} n(s, t) + \dfrac{\partial\ }{\partial s} (g(s) n(s, t))  = \dfrac{\partial^2}{\partial s^2} (D(s) n(s, t)) + \mathcal{F}[n(s, t)],
\end{equation}
where $n(s, t)$ represents the cell number density of cells of size $s$, $g(s)$ is the growth rate, $D(s)$ is the dispersion coefficient, and $\mathcal{F}$ is the fragmentation operator. Studies have examined the well-posedness and steady-state solution of equations in various forms \cite{Bernard2020Asynchronous,hall1989functional,tchouanti2024well,wake2000functional}. Zaidi et al. \cite{zaidi2014model} extended this framework to include asymmetric division, yielding the equation:
\begin{equation}
\label{eq:1.2}
\dfrac{\partial\ }{\partial t} n(s, t) + \dfrac{\partial\ }{\partial s} (g(s) n(s, t)) = \int_s^\infty b(\xi) W(s, \xi) n(\xi, t)\mathrm{d} \xi - (b(s) + \mu) n(s, t),
\end{equation}
where a mother cell of size $\xi$ divides into two daughter cells of different sizes. The steady-state size distribution for this non-local differential equation was determined using a double Dirichlet series \cite{zaidi2014model}. More recently, Xia et al. \cite{xia2020PDE} introduced unified adder-timer models, represented by the following first-order PDE:
\begin{equation}
\label{eq:1.3}
\dfrac{\partial\ }{\partial t} n(s, \varsigma, t) + (\dfrac{\partial\ }{\partial s} + \dfrac{\partial\ }{\partial \varsigma})(g(s, \varsigma, t)n(s, \varsigma, t)) = - \beta(s, \varsigma, t) (s, \varsigma, t),
\end{equation}
where $s$ represents cell size, $\varsigma$ represents the volume added since birth, and $\beta(s, \varsigma, t)$ is the division rate. This study explored the properties of different adder processes in cellular proliferation and proved the existence and uniqueness of weak solutions. 

Despite significant advances, a unified framework linking cell growth, cell cycle dynamics, cell division, and size control remains elusive. 

Cell size dynamics following divisions have also been modeled as a Markov process. This approach has been used to study the influence of noise on cell size control through random dynamical models \cite{Modi:2017aa,Teimouri:2020aa}. Nieto et al. \cite{nieto2021continuous} applied the Chepman-Kolmogorov formalism to describe the successive growth and division events, analyzing the impact of noise on size distribution dynamics. Jia et al. \cite{jia2021cell,jia2022Characterizing} developed piecewise deterministic Markov models to investigate the sizer, timer, and adder mechanisms across generations. Their findings provide insights into how size distribution is shaped by parameters governing cell cycle dynamics and experimental protocols for cell tracking. These studies raise an interesting question: how is cell size homeostasis maintained despite the inherent randomness in cellular processes?

Building on these existing models \cite{Bell1967Cell,Gabriel2019Steady,Philipp2018Analysis,xia2020PDE}, this study develops an integrated model of cell size control that incorporates cell cycle dynamics and size regulation mechanisms. The model is simplified to focus on three primary mechanisms: sizer, timer, and adder. Analytical solutions to the governing equations are derived using the method of characteristics, and steady-state size distributions are determined through numerical calculations and stochastic simulations. We investigate how factors such as growth rate and division probability influence size distributions.  Additionally, stochastic simulations are used to examine how variability in division size, modeled as a soft control strategy, affects cell size control. Finally, the model study is applied to experimental data for \textit{E. coli}.

\section{Mathematical formulations}

\begin{figure}[tbhp]
	\centering
	\includegraphics[width=10cm]{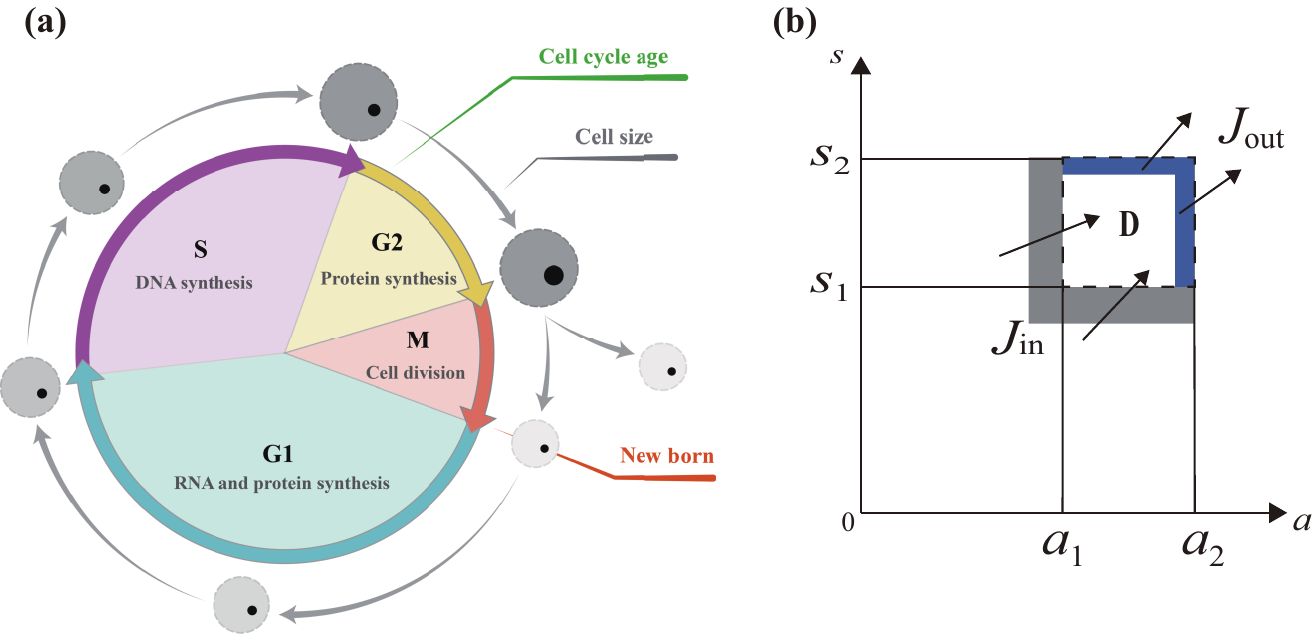}
	\caption{\textbf{Illustration of the cell cycle model and cell size evolution.} \textbf{(a)} Schematic of the cell cycle. \textbf{(b)} Flux of cell number changes: grey and blue regions represent the number of cells entering and exiting region $D$ in the next time step, respectively.}
	\label{fig:cycle-flux}
\end{figure}

The transition between cell growth and division is fundamental to maintaining cell size homeostasis. During growth, cell size typically increases, while during mitosis, a cell divides into two daughter cells, each approximately half the size of the mother cell. This study investigates the dynamic evolution of cell size within a population by developing a mathematical model that describes the processes of cell growth and division (Fig. \ref{fig:cycle-flux}a). Below, we present the mathematical framework for modeling this process.

\subsection{Cell cycle model}
\label{sec:2.1}

Consider a population of cells. Let $n(s, a, t)$ denote the number density of cells with size $s$ and cell cycle age $a$ at time $t$. To describe the temporal variation of cell numbers, we examine the changes within a small region $(s, a) \in D= [s_1, s_2]\times [a_1, a_2]$ in the size-age space over a time interval $[t, t+\Delta t]$. The variation in the cell number is expressed as:
\begin{equation}
	\Delta n(s, a, t) = \iint\limits_{D} n(s, a, t+\Delta t) \mathrm{d} \sigma - \iint\limits_{D} n(s, a, t) \mathrm{d}\sigma.
\end{equation}

The change in cell numbers within region $D$ arises from four primary components: 
\begin{enumerate}
\item[(1)] Incoming flux ($J_{\mathrm{in}}$): Cells entering $D$ from neighboring regions with smaller size or age.
\item[(2)] Outgoing flux ($J_{\mathrm{out}}$): Cells leaving $D$ for neighboring regions. 
\item[(3)] Reduction due to cell division ($J_{\mathrm{divid}}$): Cells dividing within $D$.
\item[(4)] Reduction due to cell death or removal ($J_{\mathrm{remove}}$): Cells lost from $D$ due to death or removal. 
\end{enumerate}

The incoming flux $J_{\mathrm{in}}$, corresponding to cells entering $D$ from regions with smaller size or age, is illustrated in the grey-shaded area in Fig. \ref{fig:cycle-flux}b. It is expressed as:
\begin{equation}
\begin{aligned}
	J_{\mathrm{in}} &= \int_{s_1}^{s_2} \int_{a_1 - \Delta a}^{a_1} n(s, a, t) \mathrm{d} a \mathrm{d} s + \int_{s_1-\Delta s}^{s_1} \int_{a_1}^{a_2} n(s, a, t) \mathrm{d} a \mathrm{d} s \\
 &\qquad{} + \int_{s_1-\Delta s}^{s_1} \int_{a_1 - \Delta a}^{a_1} n(s, a, t) \mathrm{d} a \mathrm{d} s  \\
	&\approx  \int_{s_1}^{s_2} n(s, a_1, t) \Delta a \mathrm{d} s  + \int_{a_1}^{a_2} n(s_1, a, t) \Delta s \mathrm{d} a  +  o(\Delta t) \\
	&\approx \Delta t\left(\int_{s_1}^{s_2} n(s, a_1, t) \mathrm{d} s + \int_{a_1}^{a_2} \left(v(s_1, a)n(s_1, a, t)\right) \mathrm{d} a \right),
\end{aligned}
\end{equation}
where $o(\Delta t)$ represents higher-order infinitesimals of $\Delta t$, $\Delta a = \Delta t$, and 
$$v(s_1, a) = \left.\frac{\Delta s}{\Delta a}\right|_{(s, a) = (s_1, a)}$$ 
is the cell size growth rate at size $s_1$ and age $a$. 

Similarly, the outgoing flux $J_{\mathrm{out}}$, representing cells leaving $D$, is given by:
\begin{equation}
J_{\mathrm{out}} \approx \Delta t\left(\int_{s_1}^{s_2} n(s, a_2, t) \mathrm{d} s + \int_{a_1}^{a_2} \left(v(s_2, a) n(s_2, a, t)\right) \mathrm{d} a \right).
\end{equation}

The reduction in cell number within $D$ due to cell division is modeled as:
\begin{equation}
J_{\mathrm{divid}} = \Delta t \iint\limits_D {\phi}(s, a) n(s, a, t) \mathrm{d} \sigma ,
\end{equation}
where $\phi(s, a)$ represents the division rate of cells with size $s$ and age $a$.

Finally, due to cell death or removal, the cell number decreases by:
\begin{equation}
J_{\mathrm{remove}} = \Delta t\iint\limits_D {\mu}(s, a) n(s, a, t) \mathrm{d} \sigma,
\end{equation}
where $\mu(s, a)$ represents the rate of cell death or removal for cells with size $s$ and cell cycle age $a$.

By combining all the components described above, the total change in the number of cells within region $D$ is given by:
\begin{equation}
\label{eq:change_n}
\begin{aligned}
	\iint\limits_{D} (n(s, a, t+\Delta t)  -  n(s, a, t)) \mathrm{d}\sigma &= J_{\mathrm{in}} - J_{\mathrm{out}} - J_{\mathrm{divid}} - J_{\mathrm{remove}}\\
	 &= -\Delta t \iint\limits_D \dfrac{\partial n(s, a, t)}{\partial a}\mathrm{d} \sigma\\
	 &\qquad{}  - \Delta t \iint\limits_D \dfrac{\partial \left(v(s, a) n(s, a, t)\right)}{\partial s} \mathrm{d} \sigma  \\
 &\qquad{}- \Delta t \iint\limits_D ({\phi}(s, a) + {\mu}(s, a)) n(s, a, t) \mathrm{d} \sigma.
 \end{aligned}
 \end{equation}
Dividing both sides of \eqref{eq:change_n} by $\Delta t$ and letting $\Delta t \to 0$, we obtain
\begin{equation}
\label{eq:intn}
\begin{aligned}
\iint_D	\dfrac{\partial n(s,a,t)}{\partial t} \mathrm{d}\sigma &=  -\iint_D \dfrac{\partial n(s,a,t)}{\partial a} \mathrm{d}\sigma - \iint_D \dfrac{\partial \left(v(s, a) n(s,a,t)\right)}{\partial s} \mathrm{d}\sigma \\
&\qquad{}
	- \iint_D ({\phi}(s, a) + {\mu}(s, a)) n(s,a,t) \mathrm{d}\sigma.
	\end{aligned}
\end{equation}

Since esquation \ref{eq:intn} holds for any region $D\in [0, +\infty) \times [0, +\infty)$, it leads to the following partial differential equation: 
\begin{equation}
	\label{eq:1}
	\dfrac{\partial n}{\partial t} + \dfrac{\partial n}{\partial a} + \dfrac{\partial \left(v(s, a) n\right)}{\partial s} = 
	- (\phi(s, a) + {\mu}(s, a)) n, \quad  a > 0, s > 0,
\end{equation}
where $n = n(s, a, t)$ represents the number density of cells at size $s$, cell cycle age $a$, and time $t$. The terms $v(s, a)$, $\phi(s, a)$, and $\mu(s, a)$ denote the cell growth rate, division rate, and cell death or removal rate, respectively. 

Let $f(s, a, t)$ denote the normalized cell size distribution, defined as:
\begin{equation}
\label{eq:f0}
	f(s, a, t) = \dfrac{n(s, a, t)}{N(t)},
\end{equation}
where $N(t)$ is the total number of cells at time $t$, expressed as: 
\begin{equation}
	N(t) = \int_0^{+\infty} \int_0^{+ \infty} {n}(s, a, t)\mathrm{d}a\mathrm{d}s.
\end{equation}

When cell death is ignored, the normalized distribution function satisfies the following equation:
\begin{equation}
	\label{eq:s2}
	\dfrac{\partial {f}(s, a, t)}{\partial t} + \dfrac{\partial {f}(s, a, t)}{\partial a} + \dfrac{\partial \left({v}(s, a){f}(s, a, t)\right)}{\partial s} = - \left(\alpha(t) + \phi(s, a)\right) {f}(s, a, t), 
\end{equation}
where $\alpha(t) = \frac{1}{N(t)}\frac{\mathrm{d}N(t)}{\mathrm{d}t}$ represents the growth rate of the cell population, which accounts for the net increase in cell numbers. 

Using the definition of $\alpha(t)$ and equation \eqref{eq:s2}, the expression for $\alpha(t)$ is derived as:
\begin{equation}
\label{eq:alpha}
\begin{aligned}
	\alpha(t) &= \dfrac{1}{N(t)}\dfrac{\mathrm{d}N(t)}{\mathrm{d}t} = \int_0^{+\infty}\int_0^{+\infty} \left(\dfrac{\partial {f}(s, a, t)}{\partial t} + \alpha(t) {f}(s, a, t)\right)\mathrm{d}s\mathrm{d}a\\
	& = -\int_0^{+\infty}\int_0^{+\infty}\left(\dfrac{\partial f(s, a, t)}{\partial a} + \dfrac{\partial (v(s, a)f(s, a, t))}{\partial s} + \phi(s, a) f(s, a, t)\right) \mathrm{d} s \mathrm{d} a.
	\end{aligned}
\end{equation}
Thus, the population growth rate $\alpha(t)$ is directly related to the division rate $\phi(s, a)$.

Equations \eqref{eq:s2}-\eqref{eq:alpha} provide a unified mathematical framework for the evolution of cell size distribution under different cell size control mechanisms. The division rate $\phi(s, a)$ is determined by the specific control mechanism and will be discussed in detail below. Furthermore, the boundary condition at $a = 0$ is also influenced by the cell size control mechanism. In this study, we focus on three simple size control mechanisms: sizer, timer, and adder.

\subsection{Sizer mechanism}
The sizer mechanism assumes that cell division occurs when a cell's size reaches a fixed threshold. Mathematically, this implies that the division rate function $\phi(s, a)$ is represented by a Dirac delta function. Let $s_d$ denote the threshold of division size. The division rate $\phi(s, a)$ is non-zero only when $s = s_d$, and is expressed as $\phi(s, a) = \delta(s - s_d)$. 

Under the sizer mechanism, neglecting cell death, the cell size distribution function satisfies the following partial differential equation:
\begin{equation}
	\label{eq:sizer1}
	\dfrac{\partial f(s, a, t)}{\partial t} + \dfrac{\partial f(s, a, t)}{\partial a} + \dfrac{\partial \left(v(s, a) f(s, a, t)\right)}{\partial s} = - \alpha(t) f(s, a, t),
\end{equation}
where $0 < s < s_d, a>0, t>0$, and $\alpha(t)$ represents the population growth rate.

The boundary condition $f(s, 0, t)$ represents the distribution of cell size $s$ produced by division. We introduce the function $p(s, s')$, which represents the division probability density that a mother cell of size $s'$ produces a daughter cell of size $s$. The probability function $p(s, s')$ satisfies the normalization condition 
\begin{equation}
\int_0^{s_d} p(s, s')\mathrm{d}s = 1, \quad \forall s' \in [0, s_d].
\end{equation}

To derive the boundary condition at $a  = 0$, we note that when $\Delta t$ is sufficiently small, mother cells with size $s_d - v(s_d, a)\Delta t < s<s_d$ will divide during the time interval $[t, t + \Delta t]$. The total number of cells dividing during this interval is: 
\begin{equation}
\label{eq:n1}
\Delta t\int_0^{+\infty} 
\int_{s_d - v(s_d, a) \Delta t}^{s_d} v(s', a) n(s', a, t) \mathrm{d} s' \mathrm{d} a.
\end{equation}
Each cell dividing at $s = s_d$ produces two daughter cells, with the size $s$ of each daughter cell distributed according to $p(s, s_d)$. Thus, the number of new birth cells with size $s$ during $[t, t+\Delta t]$ is:
$$
2 p(s, s_d) \Delta t \int_0^{+\infty} 
\int_{s_d - v(s_d, a) \Delta t}^{s_d} v(s', a) n(s', a, t) \mathrm{d} s' \mathrm{d} a.
$$
The factor $2$ accounts for the fact that each mother cell produces two daughter cells. Assuming symmetric division, the two daughter cells have equal probabilities of starting sizes.
 
Since the new birth cells have cell cycle ages $a\in [0, \Delta t]$, the number of new birth cells at time $t+\Delta t$ is given by:
\begin{equation}
\label{eq:n2}
\int_0^{\Delta t} n(s, a, t+\Delta t) \mathrm{d} a.
\end{equation}
Equating the two expressions, we have
\begin{equation}
\label{eq:n3}
\int_0^{\Delta t} n(s, a, t+\Delta t) \mathrm{d} a = 2p(s, s_d) \Delta t \int_0^{+\infty} 
\int_{s_d - v(s_d, a) \Delta t}^{s_d}  v(s', a) n(s', a, t) \mathrm{d} s' \mathrm{d} a.
\end{equation}

Letting $\Delta t \to 0$, we obtain the boundary condition:
\begin{equation}
n(s, 0, t) = 2 p(s, s_d) \int_0^{+\infty}  v(s_d, a) n(s_d, a, t) \mathrm{d} a.
\end{equation}
By normalizing $n(s, a, t)$ to $f(s, a, t)$, we obtain
\begin{equation}
\label{eq:s1}
f(s, 0, t) = 2p(s, s_d)\int_0^{+\infty} v(s_d,a)f(s_d, a, t)\mathrm{d}a. 
\end{equation}

Since no cell can have a size $s = 0$ or age $a = +\infty$, the boundary conditions at $s = 0$ or $a = +\infty$ are given as:
\begin{align}
\label{eq:bs}
f(0, a, t)  &= 0,\ \forall a \geq 0,\\
\label{eq:bainf}
\lim_{a\to +\infty}f(s, a, t) &= 0,\ \forall 0 < s < s_d.
\end{align}

Using equations \eqref{eq:s1}-\eqref{eq:bainf}, the population growth rate $\alpha(t)$ can be computed as: 
\begin{eqnarray*}
	\alpha(t) &=& -\int_0^{+\infty}\int_0^{s_d} \left(\dfrac{\partial {f}(s, a, t)}{\partial a} + \dfrac{\partial \left({v}(s, a){f}(s, a, t)\right)}{\partial s} \right)\mathrm{d}s\mathrm{d}a\\
	&=&\int_0^{s_d} f(s, 0, t) \mathrm{d} s - \int_0^{+\infty} v(s_d, a) f(s_d, a, t) \mathrm{d} a.
\end{eqnarray*}
Substituting \eqref{eq:s1}, the growth rate becomes
\begin{equation}
\alpha(t) = \int_0^{s_d}v(s_d, a)f(s_d, a, t)\mathrm{d}a.
\end{equation}

Thus, for the sizer mechanism, the governing equation and boundary conditions are:
\begin{equation}
\label{eq:sizer}
\left\{
\begin{aligned}
&\dfrac{\partial f(s, a, t)}{\partial t} + \dfrac{\partial f(s, a, t)}{\partial a} + \dfrac{\partial \left(v(s, a)f(s, a, t)\right)}{\partial s} = - \alpha(t)  f(s, a, t),\\
&\alpha(t) = \int_0^{+\infty}v(s_d, a)f(s_d, a, t)\mathrm{d}a,\\
&f(s, 0, t) = 2p(s, s_d)\displaystyle \int_0^{+\infty} v(s_d, a)f(s_d, a, t)\mathrm{d}a,\\
&f(0, a, t) = f(s, +\infty, t) = 0.
\end{aligned}
\right.
\end{equation}
The first equation is valid within the region $0 < s < s_d, a > 0, t>0$.

By integrating the governing equation over $a$, and let
$$
\tilde{f}(s, t) = \int_0^{+\infty} f(s, a, t)\mathrm{d} a
$$
represent the cell size density function, which ignores the cell cycle age, we obtain:
\begin{equation}
\label{eq:s20}
\begin{aligned}
&\dfrac{\partial \tilde{f}(s, t)}{\partial t} + \dfrac{\partial\ }{\partial s} \int_0^\infty v(s, a) f(s, a, t)\mathrm{d} a \\
= &- \alpha(t) \tilde{f}(s, t) + 2 p(s, s_d)\int_0^{+\infty} v(s_d, a) f(s_d, a, t)\mathrm{d} a.
\end{aligned}
\end{equation}

If the cell growth rate $v(s, a)$ is independent of the cell cycle age $a$, i.e., $v(s, a) = v(s)$, equation \eqref{eq:s20} simplifies to:
\begin{equation}
\label{eq:s16}
\dfrac{\partial \tilde{f}(s, t)}{\partial t} + \dfrac{\partial (v(s) \tilde{f}(s, t)) }{\partial s} = - \alpha(t) \tilde{f}(s, t) + 2 p(s, s_d) v(s_d) \tilde{f}(s_d, t),\ 0<s<s_d.
\end{equation}
Equation \eqref{eq:s16} aligns with the formulation described in \cite{Miotto2024A}.

\subsection{Timer mechanism}
The timer mechanism assumes that cell growth time is fixed, denoted by $T$. Consequently, the division rate $\phi(s, a)$ is represented by the Dirac delta function $ \phi(s, a) = \delta(a - T)$. Under this mechanism, equation \eqref{eq:s2} is simplifies to
\begin{equation}
	\label{eq:timer1}
\dfrac{\partial f(s, a, t)}{\partial t} + \dfrac {\partial f(s, a, t)}{\partial a} + \dfrac{\partial \left(v(s, a) f(s, a, t)\right)}{\partial s} 	= - \alpha(t)  f(s, a, t), 
\end{equation}
where $0 < a < T, s > 0, t > 0$, and $\alpha(t)$ represents the population growth rate. 

Unlike the sizer mechanism, the boundary condition for the timer mechanism depends on the cell cycle age $a$. Similar to the sizer mechanism, the boundary condition is expressed as:
\begin{equation}
\label{eq:timer2}
\begin{aligned}
f(s, 0, t) &= 2\int_0^T \int_0^{+\infty} p(s, s')  {\phi}(s', a)  {f}(s', a, t) \mathrm{d} s' \mathrm{d} a
\\ 	
&= 2\int_0^{+\infty} p(s, s') {f}(s', T, t)\mathrm{d}s',
\end{aligned}
\end{equation}
where $p(s, s')$ represents the probability that a mother cell of size $s'$ produces a daughter cell of size $s$.

Using the definition of $\alpha(t) $ and the boundary condition in \eqref{eq:timer2}, the population growth rate is given by:
\begin{equation}
\alpha(t) = \int_0^{+\infty} f(s, T, t)\mathrm{d}s.
\end{equation}

Thus, under the timer mechanism, the governing equation and boundary conditions are:
\begin{equation}
\label{eq:timerequation}
\left\{
\begin{aligned}
&\dfrac{\partial {f}(s, a, t)}{\partial t} + \dfrac{\partial {f}(s, a, t)}{\partial a} + \dfrac{\partial \left(v(s, a) f(s, a, t)\right)}{\partial s} =  - \alpha(t) f(s, a, t),\\
&\alpha(t) = \int_0^{+\infty} f(s, T, t)\mathrm{d}s,\\
&f(s, 0, t) = 2 \displaystyle \int_s^{+ \infty } p(s, s') {f}(s', T, t) \mathrm{d}s',\\
&f(0, a, t) = f(+\infty, a, t) = 0.
\end{aligned}
\right.
\end{equation}
Here, the first equation is valid within the region $0 < a < T, s > 0, t > 0$.

\subsection{Adder mechanism}
The adder mechanism assumes that the difference between the cell division size and its birth size is a fixed value, denoted by $\Delta s$. In this case, it is reasonable to assume that the cells can sense the size added since birth, denoted as $\varsigma = s - s_b$, where $s_b$ is the cell size at cell birth. 

We assume that the cell growth rate $v$ depends on both the cell size $s$ and the size added $\varsigma$, i.e., $v = v(s, \varsigma)$. Noting that $s = \varsigma + s_b$, we have:
\begin{equation}
\label{eq:sa}
\left\{
\begin{aligned}
&\dfrac{\mathrm{d} \varsigma}{\mathrm{d} a} = v(\varsigma + s_b, \varsigma),\\
&\varsigma(0) = 0.
\end{aligned}
\right.
\end{equation}
Let $\varsigma(a; s_b)$ be the solution of \eqref{eq:sa}. Since $v(s, \varsigma)\geq 0$, $\varsigma(a; s_b)$ is an increasing function of $a$ and hence is invertible. Let $a = \tau(\varsigma; s_b)$ denote the inverse function of $\varsigma(a; s_b)$. The cell cycle age $a$ can then be expressed in terms of the size added $\varsigma$. The cell growth time for a cell with birth size $s_b$ is given by $T = \tau(\Delta s; s_b)$. Hereafter, we write $\tau(s)$ for $\tau(s; s_b)$ for short, and hence $T = \tau(\Delta s)$.

Let $n(s, \varsigma, t \vert s_b)$ represent the number of cells with size $s$, size added $\varsigma$, at time $t$, under the condition with birth size $s_b$. From \eqref{eq:sa}, the time $a = \tau(s - s_b)$ is required for cells with birth size $s_b$ to reach size $s$. Let $f_0(s_b, t)$ represent the probability density of the birth size $s_b$ at time $t$. The number of cells with size $s$, size added $\varsigma$, at time $t$ is expressed as:
\begin{equation}
n(s, \varsigma, t) = \int_0^{+\infty} n(s, \varsigma, t \vert s_b) f_0(s_b, t - \tau(s - s_b))\mathrm{d} s_b.
\end{equation}
The corresponding cell size distribution $f(s, \varsigma, t)$ can be defined similarly to \eqref{eq:f0}:
\begin{equation}
\label{eq:addfN}
f(s, \varsigma, t) = \dfrac{n(s, \varsigma, t)}{N(t)},\quad N(t) = \int_0^{+\infty}\int_{\varsigma}^{+\infty} n(s, \varsigma, t) \mathrm{d} s \mathrm{d}\varsigma.
\end{equation}   
We note $s \geq \varsigma$.

Following a derivation similar to \eqref{sec:2.1}, we obtain the governing equation for $n(s, \varsigma, t \vert s_b)$ when $a < \tau(\Delta s)$:
\begin{equation}
\label{eq:adder1}
\dfrac{\partial n(s, \varsigma, t \vert s_b)}{\partial t} + (\dfrac{\partial\ }{\partial s} + \dfrac{\partial\ }{\partial \varsigma}) \left(v(s, \varsigma) n(s, \varsigma, t \vert s_b)\right) =  0.
\end{equation}
Integrating \eqref{eq:adder1} with $s_b$ following the density function $f_0(s_b, t - \tau(s - s_b))$, we have:
\begin{equation}
\label{eq:adder2}
\dfrac{\partial n(s, \varsigma, t)}{\partial t} + (\dfrac{\partial\ }{\partial s} + \dfrac{\partial\ }{\partial \varsigma}) (v(s, \varsigma) n(s, \varsigma, t)) = 0.
\end{equation}

From \eqref{eq:addfN} and \eqref{eq:adder2}, the cell size distribution $f(s, \varsigma, t)$ satisfies:
\begin{equation}
\dfrac{\partial f(s, \varsigma, t)}{\partial t} + (\dfrac{\partial\ }{\partial s} + \dfrac{\partial\ }{\partial \varsigma})(v(s, \varsigma)f(s, \varsigma, t)) = - \alpha(t) f(s, \varsigma, t).
\end{equation}
where
$$
\alpha(t) = \dfrac{1}{N(t)}\dfrac{\mathrm{d} N(t)}{\mathrm{d} t}.
$$

To derive the boundary condition at $\varsigma = 0$, note that all mother cells with $s(0) = s_b$ and $\varsigma  = \Delta s$ divide into two daughter cells. Thus:
\begin{equation}
\label{eq:addn0}
n(s, 0, t) = 2\int_0^{+\infty} p(s, s_b + \Delta s)  n(s_b + \Delta s, \Delta s, t \vert s_b) f_0(s_b,t - \tau(\Delta s))\mathrm{d} s_b.
\end{equation}
From \eqref{eq:addn0}, the probability density of initial cell size $s_b$ at time $t$, $f_0(s, t)$, satisfies:
\begin{equation}
\label{eq:addf0}
\begin{aligned}
f_0(s, t) &= \dfrac{n(s, 0, t)}{\int_0^{+\infty} n(s, 0, t) \mathrm{d} s}\\
& = \dfrac{n(s, 0, t)}{2 \int_0^{+\infty} n(s_b + \Delta s, \tau(\Delta s), t \vert s_b) f_0(s_0, t-\tau(\Delta s))\mathrm{d} s_b}.
\end{aligned}
\end{equation}

From \eqref{eq:addn0}, the boundary condition for $f(s, 0, t)$ is calculated as:
\begin{eqnarray*}
f(s, 0, t) &=& \dfrac{2}{N(t)}\int_0^{+\infty} p(s, s_b + \Delta s) n(s_b + \Delta s, \Delta s, t \vert s_b) f_0(s_b, t-\tau(\Delta s))\mathrm{d} s_b\\
&=& 2 \int_0^{+\infty} p(s, s_b + \Delta s) \dfrac{n(s_b + \Delta s, \Delta s, t \vert s_b)}{n(s_b+ \Delta s, \Delta s, t)} f_0(s_b, t - \tau(\Delta s)) \dfrac{n(s_b+ \Delta s, \Delta s, t)}{N(t)}\mathrm{d} s_b\\
&=&2 \int_{\Delta s}^{+\infty} p(s, s') \dfrac{n(s', \Delta s, t \vert s' - \Delta s)}{n(s', \Delta s, t)} f_0(s' - \Delta s, t - \tau(\Delta s)) \dfrac{n(s', \Delta s, t)}{N(t)}\mathrm{d} s'\\
&=&2 \int_{\Delta s}^{+\infty} p(s, s') \dfrac{n(s', \Delta s, t \vert s' - \Delta s)}{n(s', \Delta s, t)} f_0(s' - \Delta s, t - \tau(\Delta s)) f(s', \Delta s, t)\mathrm{d} s'
\end{eqnarray*}
Thus, let
$$
\varphi(s', \Delta s, t) = \dfrac{n(s', \Delta s, t \vert s' - \Delta s)}{n(s', \Delta s, t)} f_0(s' - \Delta s, t - \tau(\Delta s)),
$$
the boundary condition is given by:
\begin{equation}
f(s, 0, t) =  2 \int_{\Delta s}^{+\infty} p(s, s')\varphi(s', \Delta s, t) f(s', \Delta s, t) \mathrm{d} s'.
\end{equation}
The function $\varphi(s', \Delta s, t)$ represents the probability density of a cell with size $s'$, at time $t$, having added size $\Delta s$.

Accordingly, the population growth rate is
\begin{equation}
\begin{aligned}
\alpha(t) &= \int_0^{+\infty} v(s, 0) f(s, 0, t) \mathrm{d} s\\
&= 2 \int_0^{+\infty} \int_{\Delta s}^{+\infty} v(s, 0)p(s', s)\varphi(s', \Delta s, t) f(s', \Delta s, t) \mathrm{d} s'\mathrm{d}s.
\end{aligned}
\end{equation}

In summary, the governing equation and boundary condition under the adder mechanism are:
\begin{equation}
	\label{eq:s21}
	\left\{
	\begin{aligned}
	&\dfrac{\partial  f(s, \varsigma, t)}{\partial t}+ (\dfrac{\partial\ }{\partial s} + \dfrac{\partial\ }{\partial \varsigma}) \left(v(s, \varsigma)f(s, \varsigma, t)\right) 
	 = -\alpha(t)  f(s, \varsigma, t), \vspace{1mm}\\
	 & \alpha(t) = 2 \int_0^{\infty} \int_{\Delta s}^{+\infty} v(s, 0)p(s, s')\varphi(s', \Delta s, t) f(s', \Delta s, t) \mathrm{d} s'\mathrm{d}s,\vspace{1mm}\\
	&f(s, 0, t) = 2 \int_{\Delta s}^{+\infty} p(s, s') \varphi(s', \Delta s, t)  f(s', \Delta s, t)\mathrm{d} s', \vspace{1mm}\\
	&f(0, \varsigma, t) = f(+\infty, \varsigma, t) = 0.
	\end{aligned}
	\right.
\end{equation}
Here, the first equation is valid for $t > 0, s > 0, 0 < \varsigma < \Delta s$. This formulation aligns with the governing equation proposed by Xia et al. \cite{xia2020PDE}.

If the cell growth rate $v$ depends only on the cell cycle age, i.e., $v = v(a)$, $\varsigma(a)$ satisfies the following equation:
\begin{equation}
\left\{
\begin{aligned}
&\dfrac{\mathrm{d} \varsigma}{\mathrm{d} a} = v(a)\\
&\varsigma(0) = 0.
\end{aligned}
\right.
\end{equation}
This equation has a unique solution $\varsigma(a)$. The condition $\varsigma(T) = \Delta s$ determines the cell growth time $T$ before cell division. Therefore, the adder mechanism becomes equivalent to the timer mechanism when the cell growth rate $v$ depends solely on the cell cycle age.

\subsection{Numerical scheme}

We have derived governing equations for the sizer, timer, and adder mechanisms. The equations are first-order PDEs with boundary conditions at $a=0$ (or $\varsigma = 0$) depending on values at other boundaries ($s = s_d$, $a = T$, or $\varsigma = \Delta s$). Solving these equations numerically is non-trivial. 

In the Results section, we derive exact solutions using the method of characteristics. Additionally, we simulate the process of cell growth and cell division under these mechanisms using an individual-cell-based model implemented via stochastic simulation in \verb|C++|. The simulation tracks each cell's size and age over time, determining cell fates (growth or division) based on the control mechanism. The Algorithm below summarizes the stochastic simulation scheme.

\begin{algorithm}
\label{alg:01}
\caption{Stochastic simulation for cell growth and division}
\begin{algorithmic}[1]
\STATE{\textbf{Input:} Control functions: cell growth rate (\verb|function| $v$), cell division condition (\verb|function| $\phi(s, a)$), probability of daughter cell size (\verb|function| $p(s, s')$); control parameters: $s_d$ (sizer), $T$ (timer), or $\Delta s$ (adder); time step $\Delta t$; simulation end time $T_{\mathrm{end}}$.}
\STATE{\textbf{Initialize:} Set initial time $t = 0$ and initial cell number $N_0$. Randomly assign initial cycle ages and sizes to all cells. }
\FOR {$t=0$ to $T_{\mathrm{end}}$ with step $\Delta t$ }
\STATE{\textbf{Step 1: Cell fate decision}}
\FOR {each cell }
\STATE{ Evaluate whether the cell satisfies the division condition $\phi(s, a)$ during the time interval $(t, t + \Delta t)$}. 
\STATE{If the division condition is satisfied, the cell divides into two daughter cells. Otherwise, the cell continues to grow at the rate $v$.}
\ENDFOR
\STATE{\textbf{Step 2: System update}} 
\FOR{ each cell }
\IF{the cell undergoes division (mitosis)}
\STATE{Generate two daughter cells. Assign sizes to the daughter cells based on the probability density function $p(s, s')$. Set the cycle age for each daughter cell as $a = 0$. Update the total cell count: $N = N+1$.}
\ELSE
\STATE{If the cell remains in the growth phase, update its cycle age: $a = a + \Delta t$. The total cell count $N$ remains unchanged.}
\ENDIF
\ENDFOR
\ENDFOR
\end{algorithmic}
\end{algorithm}

During stochastic simulation, we note that the number of cells can increase significantly due to cell division, posing a significant challenge for data storage. To address this issue, we employed a downsampling strategy to limit the number of simulated and stored cells. Specifically, we defined a maximum cell number, denoted as $N_{\max} =  5\times 10^4$. In step 2 of the system update, whenever the total number of cells $N$ exceeds $N_{\max}$, we randomly select $N_{\max}$ cells to continue in the next round of simulation. 

\section{Results}

\subsection{Collins-Richmond formula}

In the model equations established in the previous section, the cell growth rate $v$ plays a crucial role in maintaining the balance between cell growth and division. However, the cell growth rate cannot be measured directly, making it important to derive the dependence of cell growth rate on size from available experimental data. 
 
Experimentally, the cell size distribution can be obtained by measuring the sizes of individual cells. However, it is difficult to measure the cell cycle age directly. Therefore, the experimentally obtained cell size distribution is often a marginal distribution that disregards the cell cycle age, i.e., 
\begin{equation}
\tilde{f}(s, t) = \int_0^{+\infty}f(s, a, t)\mathrm{d}a.
\end{equation}

The Collins-Richmond formula in \cite{1962Rate} provides a method of calculating the cell growth rate from stable cell size distributions. However, the derivation in \cite{1962Rate} is difficult to follow. Here, we rederive the Collins-Richmond formula based on the model equations presented in \eqref{sec:2.1}.

First, integrating \eqref{eq:s2} over $a$, we obtain the equation for $\tilde{f}(s, t)$:
\begin{equation}
\label{eq:3.2}
\dfrac{\partial \tilde{f}(s, t)}{\partial t}  + \dfrac{\partial (\tilde{v}(s, t)\tilde{f}(s, t))}{\partial s} = -(\alpha(t) + \tilde{\phi}(s, t)) \tilde{f}(s, t) + f(s, 0, t),
\end{equation}
where we have applied the boundary condition $f(s, +\infty, t) = 0$, and $\tilde{v}(s, t)$ and $\tilde{\phi}(s, t)$ are defined as: 
\begin{equation*}
\begin{aligned}
\tilde{v}(s, t)\tilde{f}(s, t) &= \int_0^{+\infty}v(s, a)f(s, a, t)\mathrm{d}a,\\
\tilde{\phi}(s, t)\tilde{f}(s, t) &= \int_0^{+\infty}\phi(s, a) f(s, a, t)\mathrm{d}a.
\end{aligned}
\end{equation*}

Since cell division is the only cause for changes in cell number, the population growth rate $\alpha$ equals the fraction of cells at the mitosis phase and can be obtained experimentally. 

Let $h(s, t)$ represent the size distribution of cells during the mitosis phase. This distribution is defined as:
\begin{equation}
\tilde{h}(s, t) = \dfrac{1}{\alpha(t)}\tilde{\phi}(s, t)\tilde{f}(s, t).
\end{equation}

Moreover, since each cell divides into two daughter cells, the size distribution of newborn cells is given by
\begin{equation}
f_0(s, t) = \dfrac{1}{2 \alpha(t)} f(s, 0, t).
\end{equation}

Now, assuming a state of homeostasis where the cell size distribution does not change with time, let $\hat{f}(s)$ represent the homeostasis distribution, and $\hat{f}_0(s) = \lim_{t\to\infty} f_0(s, t)$ represent the distribution of newborn cell sizes at homeostasis. Additionally, the functions $\tilde{v}(s, t)$, $\tilde{h}(s, t)$, and $\alpha(t)$ will reach the steady-state, denoted $\hat{v}(s)$, $\hat{h}(s)$, and $\alpha$, respectively. 

From equation \eqref{eq:3.2}, the homeostatic distribution $\hat{f}(s)$ satisfies:   
\begin{equation}
\label{eq:3.3}
\dfrac{\mathrm{d}(\hat{v}(s)\hat{f}(s))}{\mathrm{d}s}  = - \alpha \hat{f}(s) - \alpha \hat{h}(s) + 2 \alpha \hat{f}_0(s).
\end{equation}

Integrating \eqref{eq:3.3} with respect to $s$, and introducing the following distribution functions:
\begin{equation}
F(s) = \int_0^s \hat{f}(s')\mathrm{d}s',\ F_0(s) = \int_0^s \hat{f}_0(s')\mathrm{d}s',\ H(s) = \int_0^s \hat{\phi}(s')\hat{f}(s')\mathrm{d}s',
\end{equation}
we obtain
\begin{equation}
\label{eq:3.7}
\hat{v}(s) \hat{f}(s) = - \alpha F(s) + 2 \alpha F_0(s)  - \alpha H(s).
\end{equation}

If the cell size increment $\Delta s$ from new birth to the next division, and let $\delta(\Delta s)$ be the distribution function of $\Delta s$, then $H(s)$ can be expressed as the convolution
\begin{equation}
H(s) = (F_0 *\delta)(s).
\end{equation}

Thus,  from equation \eqref{eq:3.7}, the cell growth rate $\hat{v}(s)$ can be obtained from the distribution functions $F(s)$, $F_0(s)$, $\hat{f}(s)$, and $\delta(s)$ at homeostasis as:
\begin{equation}
\label{eq:3.9}
\hat{v}(s) = \dfrac{1}{\hat{f}(s)}(2\alpha F_0(s) - \alpha(F_0*\delta)(s) - \alpha F(s)).
\end{equation}

Equation \eqref{eq:3.9} represents the Collins-Richmond formula, first derived in 1962 \cite{1962Rate}. This formula is crucial for estimating the dependence of cell growth rate on cell size through experimental data \cite{Tzur2009Cell}.

\subsection{Sizer mechanism}
In this section, we analyze the size distribution for the sizer mechanism. 

For simplicity, we normalize the cell size so that $s_d = 1$. Additionally, we assume that the probability density function $p(s, s')$ depends on the ratio $s/s'$ so that $p(s, 1) = p(s)$. The probability density function $p(s)$ satisfies the normalization condition:
\begin{equation}
\int_0^1 p(s) \mathrm{d} s = 1.
\end{equation}

Using these conventions, the governing equation for the sizer mechanism is written as:
\begin{equation}
	\label{eq:size3}
	\left\{
	\begin{aligned}
	&\dfrac{\partial f(s, a, t)}{\partial t}  + \dfrac{\partial f(s, a, t)}{\partial a} + \dfrac{\partial (v(s, a)f(s, a, t))}{\partial s}  = -\alpha(t) f(s, a, t),\\
	&\alpha(t) = \int_0^{+\infty} v(1, a') f(1, a', t) \mathrm{d} a',\\ 
	&f(s, 0, t) = 2 p(s) \alpha(t), \\
	&f(0, a, t) = f(s, +\infty, t) = 0.
	\end{aligned}
	\right.
\end{equation}

We are interested only in the biologically meaningful solution, which must be nonnegative and normalized as a probability density function. Hence, the following conditions must hold:
\begin{equation}
\label{eq:size4}
\left\{
\begin{aligned}
	&f(s, a, t) \geq 0,\quad 0<s<1, a > 0, t > 0,\\
	&\int_0^{+\infty}\int_0^1f(s', a', t) \mathrm{d}s'\mathrm{d} a' = 1,\forall t.\\
\end{aligned}
\right.
\end{equation}

The initial condition is assumed to be:
\begin{equation}
\label{eq:size5}
f(s, a, 0) = g(s, a),
\end{equation}
where $g(s, a)$ satisfies the normalization condition given by \eqref{eq:size4}.

\begin{lemma}
\label{le:30}
Consider equation \eqref{eq:size3}. If the initial condition \eqref{eq:size5} satifies the normalization condition 
$$
\int_0^{+\infty}\int_0^1 g(s, a) \mathrm{d} s \mathrm{d} a = 1,
$$
the solution $f(s, a, t)$ satisfies normalization condition 
$$
\int_0^{+\infty}\int_0^1 f(s, a, t) \mathrm{d} s \mathrm{d} a = 1
$$
for any $t > 0$.
\end{lemma}
\begin{proof}
Let
$$
F(t) = \int_0^{+\infty}\int_0^1 f(s, a, t) \mathrm{d} s \mathrm{d} a.
$$
Integrating \eqref{eq:size3} over $s$ and $a$, we obtain:
\begin{eqnarray*}
\dfrac{\mathrm{d} F(t)}{\mathrm{d} t} &=& - \alpha(t) F(t) - \int_0^{+\infty} v(1, a) f(1, a, t) \mathrm{d} a + \int_0^1 f(s, 0, t) \mathrm{d} s  \\
&=& - \alpha(t) F(t) - \alpha(t) + \int_0^1 2 p(s) \alpha(t) \mathrm{d} s\\
&=& \alpha(t) (1 - F(t)).
\end{eqnarray*}
Thus, we have $F(t) \equiv 1$ whenever $F(0) = 1$.
\end{proof}

In the following, we derive the solution of equations \eqref{eq:size3}-\eqref{eq:size5}, prove the stability of the steady-state solution, and analyze the dependence of size distribution on the model parameters.

\subsubsection{Exact solution}
\label{sec:1}
The exact solution of \eqref{eq:size3}-\eqref{eq:size5} can be obtained by the method of characteristics, and the result is presented in Equation Theorem \ref{th:3.1} below. 

Before we state the theorem, we introduce some necessary notations. Consider the differential equation:
\begin{equation}
\label{eq:sa0}
\dfrac{\mathrm{d} s}{\mathrm{d} a} = v(s, a).
\end{equation}
Let $u(s, a)$ be a first integral of \eqref{eq:sa0}, and define the function $s = w(\varphi, a)$ through the implicit equation 
$$\varphi = u(s, a).$$

Additionally, denote
\begin{equation}
\gamma_{a'}(s, a) = w(u(s, a), a').
\end{equation}

We now state the following theorem, which provides the exact solution to equation \eqref{eq:size3}.
\begin{theorem}
\label{th:3.1}
Let
\begin{equation}
\psi(a; \varphi_1, \varphi_2) = \alpha(a+\varphi_1) + v'_s(w(\varphi_2, a), a),
\end{equation}
where $v'_s(s, a)$ denotes the derivative of $v(s, a)$ with respect to $s$, and
$$
\alpha(t) = \int_0^{+\infty} v(1, a') f(1, a', t) \mathrm{d} a'.
$$
The solution of equation \eqref{eq:size3} is given by 
\begin{equation}
	\label{eq:schara}
	 f(s, a, t) =
	\begin{cases}
\displaystyle		2 p(\gamma_0(s, a)) \alpha(t-a) e^{-\int_{0}^a \psi(a'; t-a, u(s, a))\mathrm{d} a'} & a < t, \gamma_0(s,a) \geq 0,\vspace{1mm}\\
\displaystyle		g(\gamma_{a-t}(s, a), a - t) e^{-\int_0^t \psi(a-t'; t-a, u(s, a))\mathrm{d} t'}& a \geq t, \gamma_{a-t}(s, a) \geq 0,\vspace{1mm}\\
		0 & \text{otherwise}.
	\end{cases}
\end{equation}
Here, $0\leq s \leq 1, a \geq 0, t\geq 0$.
\end{theorem}
\begin{proof}
We rewrite the governing equation in \eqref{eq:size3} as
\begin{equation}
\label{eq:size6}
\dfrac{\partial f}{\partial t} + \dfrac{\partial f}{\partial a} + v(s, a) \dfrac{\partial f}{\partial s} = -(\alpha(t) +v'_s(s, a)) f.
\end{equation}
The characteristic equation corresponding to \eqref{eq:size6} is:
\begin{equation}
\label{eq:3.14}
\dfrac{\mathrm{d}t}{1} = \dfrac{\mathrm{d}a}{1} = \dfrac{\mathrm{d}s}{ v(s, a)} = \dfrac{\mathrm{d} f}{-(\alpha(t) +  v'_s(s, a)) f}.
\end{equation}

From \eqref{eq:3.14}, we obtain two independent first integrals:
$$
\varphi_1 = t - a, \varphi_2 = u(s, a).
$$
Moreover, we have:
$$
\begin{aligned}
\dfrac{\mathrm{d} f}{\mathrm{d} a} &= - (\alpha(a + \varphi_1) + v'_s(w(\varphi_2, a), a)) f\\
&= - \psi(a; \varphi_1, \varphi_2) f.
\end{aligned}
$$
This leads to the third first integral:
$$
\varphi_3 = f e^{\int_0^a \psi(a'; \varphi_1, \varphi_2)\mathrm{d} a'}.
$$

Using the theory of first-order partial differential equations, the implicit general solution of \eqref{eq:size6} is given by
\begin{equation}
	\Phi(\varphi_1, \varphi_2,  \varphi_3) = 0
\end{equation}
where $\Phi (\varphi_1, \varphi_2, \varphi_3)$ is a continuous function and $\Phi'_{\varphi_3} \neq 0$. Thus, the general solution $f(s, a,t)$ is expressed as
\begin{equation}
\label{eq:s25}
\begin{aligned}
f(s, a, t) &= \Psi(\varphi_1, \varphi_2)e^{-\int_0^a \psi(a'; \varphi_1, \varphi_2) \mathrm{d}a'} \vspace{1mm}\\
&= \Psi(t-a, u(s, a))e^{-\int_0^a \psi(a'; t-a, u(s, a))\mathrm{d}{a'}}.
\end{aligned}
\end{equation}
	
To determine the function $\Psi$, we consider the boundary condition at $a = 0$: 
\begin{equation}
\label{eq:s23}
f(s, 0, t) = \Psi(t, u(s, 0)) = 2 p(s) \alpha(t).
\end{equation} 
From $s = w(u(s, 0), 0)$, we have:
\begin{equation}
\Psi(t, u) =  2 p(w(u,0)) \alpha(t).
\end{equation}

Thus, when $t > a$, 
\begin{equation}
\label{eq:s30}
\begin{aligned}
f(s, a, t) &= \Psi(t-a, u(s, a)) e^{-\int_0^a \psi(a'; t-a, u(s, a))\mathrm{d}a'} \\
&= 2 p(w(u(s,a),0)) \alpha(t-a) e^{-\int_0^a \psi(a'; t-a, u(s, a))\mathrm{d}a'}\\
&= 2 p(\gamma_0(s, a)) \alpha(t-a) e^{-\int_{0}^a \psi(a'; t-a, u(s, a))\mathrm{d} a'}.
\end{aligned}
\end{equation}

Next, using the initial condition $f(s, a, 0) = g(s, a)$, we find: 
\begin{equation}
	\label{eq:s24}
	 g(s, a) = f(s, a, 0) = \Psi(-a, u(s, a)) e^{-\int_0^a \psi(a'; -a, u(s, a)) \mathrm{d} a'}.
\end{equation}
Thus, since $s = w(u(s, a), a)$, we obtain:
\begin{equation}
\Psi(-a, u) = g(w(u, a), a) e^{\int_0^a \psi(a'; -a, u)\mathrm{d}a'}
\end{equation}
for $a > 0$. 

Hence, for $ t < a$,  we have:
\begin{eqnarray*}
\Psi(t - a, u(s, a)) &=& g(w(u(s, a), a - t), a - t) e^{\int_0^{a-t} \psi(a'; t-a, u(s, a))\mathrm{d}a'}\\
 &=& g(\gamma_{a-t}(s, a), a-t) e^{\int_0^{a-t} \psi(a'; t-a, u(s, a))\mathrm{d}a'}.
\end{eqnarray*}
Thus, 
\begin{equation}
\label{eq:s31}
\begin{aligned}
f(s, a, t) &= g(\gamma_{a-t}(s, a), a-t) e^{-\int_{a-t}^a \psi(t'; t-a, u(s, a))\mathrm{d} t'}\\
&=g(\gamma_{a-t}(s, a), a-t) e^{-\int_{0}^t \psi(a-t'; t-a, u(s, a))\mathrm{d} t'}.
\end{aligned}
\end{equation}

Combining equations \eqref{eq:s30} and \eqref{eq:s31}, the solution of \eqref{eq:size3} is given by:
\begin{equation}
	 f(s, a, t) =
	\begin{cases}
\displaystyle	2 p(\gamma_0(s, a)) \alpha(t-a) e^{-\int_0^a \psi(a'; t-a, u(s, a))\mathrm{d} a'} & a < t, \gamma_0(s, a) \geq 0,\vspace{1mm}\\
\displaystyle	g(\gamma_{a-t}(s, a), a - t) e^{-\int_0^t \psi(a-t'; t-a, u(s, a))\mathrm{d} t'}& a \geq t, \gamma_{a-t}(s, a) \geq 0,\\
		0 & \text{otherwise}.
	\end{cases}
\end{equation}
Theorem \ref{th:3.1} is proved.
\end{proof}

Using \eqref{eq:schara}, it is straightforward to obtain the solution for two specific cases.

For the case where $v(s, a) = v(a)$, let 
$$u(a) = \int_0^a v(a') \mathrm{d} a',$$
then the following holds:
\begin{eqnarray*}
\psi'(a; \varphi_1, \varphi_2) &=& \alpha(a + \varphi_1),\\
\gamma_{a'}(s, a) &=& s - u(a) + u(a').
\end{eqnarray*} 
Hence, the solution given by \eqref{eq:schara} becomes:
\begin{equation}
f(s, a, t) = 
\begin{cases}
\displaystyle 2 p(s - u(a)) \alpha(t - a) e^{-\int_{t-a}^t\alpha(a')\mathrm{d} a'} & a < t, s\geq u(a),\\
\displaystyle g(s-u(a)+u(a-t), a-t) e^{-\int_0^t \alpha(t') \mathrm{d} t'} & a \geq t, s \geq u(a-t) - u(a),\\
0 & \mathrm{otherwise.}
\end{cases}
\end{equation}
In the limit as $t\to+\infty$, the steady-state solution $\hat{f}(s, a)$ is expressed as:
\begin{equation}
\begin{aligned}
\displaystyle \hat{f}(s, a) = 2 p(s - u(a)) \alpha e^{-\alpha a}\\
\displaystyle \alpha = \int_0^{+\infty} v(a') \tilde{f}(1, \alpha')\mathrm{d} a'.
\end{aligned}
\end{equation}

For the case where $v(s, a) = v(a) s$, the following relationships hold:
\begin{eqnarray*}
\psi'(a'; \varphi_1, \varphi_2) &=& \alpha(a + \varphi_1) + v(a)\\
\gamma_{a'}(s, a) &=& s e^{u(a') - u(a)}.
\end{eqnarray*}
The solution \eqref{eq:schara} is then:
\begin{equation}
f(s, a, t) = 
\begin{cases}
\displaystyle 2 p(s e^{-u(a)}) \alpha(t - a) e^{-(\int_{t-a}^t\alpha(a')\mathrm{d} a' + u(a))} & a < t, s\geq 0,\\
\displaystyle g(s e^{-(u(a) - u(a - t))}, a-t) e^{-(\int_0^t \alpha(t')\mathrm{d} t'+ u(a) - u(a-t))} & a \geq t, s \geq 0,\\
0 & \text{otherwise}.
\end{cases}
\end{equation}
In the limit $t\to+\infty$, the steady-state solution $\hat{f}(s, a)$ is expressed as:
\begin{equation}
\begin{aligned}
\displaystyle \hat{f}(s, a) = 2 p(s e^{- u(a)}) \alpha e^{-\alpha a}\\
\displaystyle \alpha = \int_0^{+\infty} v(a') \tilde{f}(1, \alpha')\mathrm{d} a'.
\end{aligned}
\end{equation}

Given the initial condition $f(s, a,  0) = g(s, a)$, equation \eqref{eq:schara} provides an iterative scheme for solving the exact solution. Moreover, the solution is uniquely determined by the initial condition. To calculate $f(s, a, t)$, we need to know $\alpha(t - a)$, for which the boundary condition $f(1, a, t')$ for $t' < t$ is required. 

\subsubsection{Existence and stability of the steady-state solution}

In experimental studies, we are often interested in the steady-state solution as $t\to +\infty$.

As $t$ approaches infinity, the net growth rate $\alpha(t)$ becomes a constant value, and the function $\psi(a'; t-a, u(s, a))$ tends to $\alpha + v'_s(w(u(s, a), a'), a')$. Therefore, the steady-state solution is given by:
\begin{equation}
\label{eq:ssf}
\hat{f}(s, a) = 
\begin{cases}
2 \alpha p(\gamma_0(s, a)) e^{-\int_0^a (\alpha + v'_s(w(u(s, a), a'), a'))\mathrm{d} a'}, & \gamma_0(s, a) \geq 0\\
0, & \text{otherwise}.
\end{cases}
\end{equation}
The population growth rate $\alpha$ is:
\begin{equation}
\label{eq:ssa}
\alpha = \int_0^{+\infty} v(1, a') \hat{f}(1, a') \mathrm{d} a'.
\end{equation}

Before stating the theorem for the stability of the steady state, let us introduce some notations. Let $s(a; s_b)$ be the solution of the differential equation 
$$
\left\{
\begin{aligned}
&\dfrac{\mathrm{d} s}{\mathrm{d} a} = v(s, a)\\
&s(0) = s_b.
\end{aligned}
\right.
$$
Define $A$ as the maximum value of $a$ such that $s(a; s_b) = 1$ for $s_b \in [0, 1]$. Since $s(a; s_b)$ is a monotonic increasing function of $a$, $A$ is well defined. Furthermore, for $a\geq A$, we have $f(s, a, t) = 0$. Thus, we can limit our analysis to the region $0\leq s \leq 1, 0\leq a \leq A$. 

We also define the following parameters:
\begin{equation}
\label{eq:ssB}
B = \sup_{0 \leq a\leq A, 0\leq s \leq 1} \left\vert\frac{v'_a(s, a)}{v(s, a)}\right\vert, 
\end{equation}
and
\begin{equation}
\label{eq:sspq}
P = \int_0^1 p(s)^2 v(s, 0) \mathrm{d} s,\ Q = \int_0^1\int_0^{A} \hat{f}(s, a)^2 v(s, a) \mathrm{d} a \mathrm{d} s.
\end{equation}

\begin{theorem}
Consider the sizer mechanism described by equation \eqref{eq:size3}, and let $\hat{f}(s, a)$ be the steady-state solution given by \eqref{eq:ssf} and \eqref{eq:ssa}. Let $\alpha, A, B, P, Q$ be the values defined above. If $2 \alpha > B$ and 
$$
\Delta = A Q -  (2 \alpha - B) (1 - 4 A P) < 0,
$$
the steady-state solution $\hat{f}(s, a)$ is asymtotically stable.  
\end{theorem}

\begin{proof}
To analyze the stability of the steady state, we introduce a small perturbation $\varepsilon e^{\lambda t} w(s, a)$ to the solution $\hat{f}(s, a)$. Thus, we assume:
\begin{equation}
 f(s, a, t) = \hat{f}(s, a) + \varepsilon e^{\lambda t} w(s, a).
\end{equation}
Substituting this into the governing equation \eqref{eq:size3}, we obtain the linearized perturbation equation:
\begin{equation}
\label{eq:perturbations}
\left\{
\begin{aligned}
&\lambda w(s, a) = -\dfrac{\partial w(s, a) }{\partial a} - \dfrac{\partial (v(s, a) w(s, a)) }{\partial s} - \alpha w(s, a)\\
&\quad\qquad\qquad{} - \hat{f}(s, a) \int_0^{+\infty} v(1, a) w(1, a) \mathrm{d} a,\\
&w(s, 0) = 2 p(s) \int_0^{+\infty} v(1, a) w(1, a) \mathrm{d} a,\\
&w(0, a) = w(s, +\infty) = 0,\\
&\int_0^{+\infty}\int_0^1 w(s, a) \mathrm{d}s \mathrm{d} a = 0.
\end{aligned}
\right.
\end{equation}
	
Multiplying both sides of \eqref{eq:perturbations} by $v(s, a) w(s, a)$, integrating over $s$ and $a$, and applying the boundary conditions, we arrive at the following equation:
\begin{eqnarray*}
\lambda \int_0^1\int_0^{+\infty} (v w^2)\mathrm{d}a\mathrm{d}s &=& -\int_0^1\int_0^{+\infty}\left[v w \dfrac{\partial w}{\partial a} + (v w) \dfrac{\partial (v w)}{\partial s} + \alpha v w^2\right]\mathrm{d}a\mathrm{d}s\\
&&{} - \left(\int_0^1 \int_0^{+\infty} (\hat{f} v w) \mathrm{d} a \mathrm{d} s\right)\times \left(\int_0^{+\infty} v(1,a) w(1, a)\mathrm{d}a\right)\\
&=& \dfrac{1}{2}\int_0^1 v(s, 0) w(s, 0)^2\mathrm{d}s + \dfrac{1}{2}\int_0^1 \int_0^{+\infty} v'_a(s, a) w(s, a)^2 \mathrm{d} a \mathrm{d} s \\
&&{} - \dfrac{1}{2}\int_0^{+\infty }  (v(1, a)w(1,a))^2 \mathrm{d}a - \alpha \int_0^1\int_0^{+\infty } (v w^2) \mathrm{d}a \mathrm{d}s\\
&&{} - \left(\int_0^1 \int_0^{+\infty} (\hat{f} v w) \mathrm{d}a \mathrm{d}s\right) \times \left(\int_0^{+\infty} v(1, a) w(1, a) \mathrm{d} a\right).
\end{eqnarray*}

Now, we estimate each term on the right-hand side of the above equation. 

Since $v(s, a) \geq 0$, we always have
$$
\int_0^1\int_0^{+\infty} (v w^2) \mathrm{d} a \mathrm{d} s \geq 0,
$$
and
$$
\int_0^{+\infty} (v(1, a) w(1, a))^2 \mathrm{d} a \geq 0.
$$
Thus, we denote
$$
X^2 = \int_0^1\int_0^{+\infty} (v w^2) \mathrm{d} a \mathrm{d} s, Y^2 = \int_0^{+\infty} (v(1, a) w(1, a))^2 \mathrm{d} a.
$$

First, 
$$
\begin{aligned}
\frac{1}{2}\int_0^1 v(s, 0) w(s, 0)^2 \mathrm{d} s &= \frac{1}{2}\left(\int_0^1 v(s, 0) (2 p(s))^2 \mathrm{d} s\right)\times \left(\int_0^{+\infty} v(1, a) w(1, a) \mathrm{d} a\right)^2\\
&= 2 P \left(\int_0^{+\infty} v(1, a) w(1, a) \mathrm{d} a\right)^2.
\end{aligned}
$$
From the definition of $A$, we have $w(s, a) =0$ whenever $a\geq A$. Hence, applying the Cauchy-Schwarz inequality, 
$$
\left(\int_0^{+\infty} v(1, a) w(1, a) \mathrm{d} a\right)^2 = \left(\int_0^{A} v(1, a) w(1, a) \mathrm{d} a\right)^2 \leq A Y^2.
$$
Thus, 
\begin{equation}
\label{eq:ss35}
\frac{1}{2}\int_0^1 v(s, 0) w(s, 0)^2 \mathrm{d} s \leq 2 A P Y^2.
\end{equation}

Next, 
\begin{equation}
\label{eq:ss36}
\begin{aligned}
&- \left(\int_0^1 \int_0^{+\infty} (\hat{f} v w) \mathrm{d}a \mathrm{d}s\right)\times \left(\int_0^{+\infty} v(1, a) w(1, a) \mathrm{d} a\right)\\
 \leq &\left\vert \int_0^1 \int_0^{+\infty} (\hat{f} v w) \mathrm{d}a \mathrm{d}s \right\vert \times \left\vert \int_0^{+\infty} v(1, a) w(1, a) \mathrm{d} a \right\vert\\
 \leq & \left(\int_0^1\int_0^{+\infty} \hat{f}^2 v \mathrm{d} a \mathrm{d} s\right)^{1/2} \left(\int_0^1 \int_0^{+\infty} v w^2 \mathrm{d}a\mathrm{d}s\right)^{1/2}\times \sqrt{A} Y\\
 =& \sqrt{A Q} X Y.
 \end{aligned}
\end{equation}

Moreover, from the definition of $B$,
\begin{equation}
\label{eq:ss37}
\dfrac{1}{2}\int_0^1\int_0^{+\infty} v'_a(s, a) w(s, a)^2 \mathrm{d}a \mathrm{d} s \leq \dfrac{1}{2}B \int_0^1\int_0^{+\infty} v w^2 \mathrm{d} a \mathrm{d} s = \dfrac{1}{2}B X^2.
\end{equation}

Thus, applying equations \eqref{eq:ss35}--\eqref{eq:ss37}, we have
\begin{eqnarray*}
\lambda X^2 &\leq& 2 PA Y^2 + \dfrac{1}{2}B X^2 - \frac{1}{2} Y^2 - \alpha X^2 + \sqrt{A Q} X Y\\
&=& - ((\alpha - \dfrac{1}{2}B)X^2 - \sqrt{A Q} X Y + (\frac{1}{2} - 2 A P)Y^2).
\end{eqnarray*}
Thus, if $2 \alpha -B > 0$ and 
$$
\Delta = A Q - (2 \alpha - B) (1 - 4 A P) < 0,
$$
we always have $\lambda X^2 < 0$, which implies $\lambda < 0$, and hence the steady-state solution is asymtotically stable. 
\end{proof}

\subsubsection{Linear growth rate}

Linear cell size growth has been reported in many studies \cite{Iyer-Biswas:2014aa,schaechter1958dependency, van2018cell, vargas2018cell}. When the cell growth rate is independent to $a$ and linearly depends on $s$, i.e., $v(s) = v_0 + v_1 s$, the following expressions hold:
\begin{eqnarray*}
u(s, a) &=& e^{-v_1 a} (s + \frac{v_0}{v_1}),\\
w(\varphi_2, a) &=& \varphi_2 e^{v_1 a} - \frac{v_0}{v_1},\\
\gamma_0(s, a) &=& s e^{-v_1 a} - \frac{v_0}{v_1} (1 - e^{-v_1 a}),\\
\gamma_{a-t}(s, a) &=& s e^{-v_1 t} - \frac{v_0}{v_1} (1 - e^{-v_1 t}),\\
\psi(a'; t-a, u(s, a)) &=& \alpha(a' + t - a) + v_1,\\
\psi(a - t'; t-a, u(s, a)) &=& \alpha(t-t') + v_1.
\end{eqnarray*}
Using these expressions, the solution given by \eqref{eq:schara} becomes:
\begin{equation}
\label{eq:sizel25}
f(s, a, t) = 
\begin{cases}
\displaystyle 2 p(\gamma_0(s, a)) \alpha(t-a)e^{-\int_{t-a}^t (\alpha(t') + v_1) \mathrm{d} t'} & a < t, s \geq  \frac{v_0}{v_1}(e^{v_1 a} - 1)\vspace{1mm}\\
\displaystyle g(\gamma_{a-t}(s, a), a-t)e^{-\int_0^t (\alpha(t') + v_1) \mathrm{d} t'} & a \geq t, s \geq \frac{v_0}{v_1} (e^{v_1 t} - 1)\vspace{1mm}\\
0 & \text{otherwise},
\end{cases}
\end{equation}
where
\begin{equation}
\alpha(t) = \int_{\gamma_0(1, a) \geq 0} v(1) f(1, a, t) \mathrm{d} a.
\end{equation}

Linear growth rates encompass both linear growth ($v_1 = 0$) and exponential growth ($v_0 = 0$), which have been observed in certain unicellular organisms. 

In the steady-state limit as $t\to +\infty$, the solution becomes:
\begin{equation}
\label{eq:ls39}
\hat{f}(s, a) = 
\begin{cases}
\displaystyle 2 \alpha p(\gamma_0(s, a)) e^{-(\alpha + v_1) a} & s > \frac{v_0}{v_1} (e^{v_1 a} - 1)\vspace{1mm}\\
\displaystyle 0 & \text{otherwise}.
\end{cases}\\
\end{equation}
The population growth rate $\alpha$ satisfies:
\begin{equation}
\label{eq:ls40}
\begin{aligned}
\alpha  &= \int_{\gamma_0(1, a) \geq 0} v(1)\hat{f}(1, a) \mathrm{d} a\\
&= \int_{\gamma_0(1, a) \geq 0} 2 \alpha v(1) p(\gamma_0(1, a)) e^{-(\alpha + v_1)a}\mathrm{d} a\\
&= 2 \alpha \int_0^{+\infty} v(1) p(e^{-v_1 a} - \frac{v_0}{v_1}(1 - e^{-v_1 a})) e^{-(\alpha + v_1)a}\mathrm{d} a\\
&= 2 \alpha \int_0^1 p(u) \left(\frac{v_0 + v_1}{v_0 + u v_1}\right)^{-\alpha/v_1}\mathrm{d} u. 
\end{aligned}
\end{equation}

Equation \eqref{eq:ls40} has a singularity when $v_1 = 0$. In this case, as $v_1 \to 0$ in \eqref{eq:ls40}, we obtain:
\begin{equation}
\alpha = 2 \alpha \int_0^1 p(u) e^{-\frac{(1-u)\alpha}{v_0}}\mathrm{d} u.
\end{equation}

To unify these cases, we define the function $K_0(\alpha, s, u)$ as:
\begin{equation}
\label{eq:Ks}
K_0(\alpha, s, u) = 
\begin{cases}
\displaystyle \left( \frac{v_0 + s v_1}{v_0 + u v_1}\right)^{-\alpha/v_1}& \mathrm{if}\ v_1 \not= 0\vspace{1mm}\\
\displaystyle e^{-\frac{(s-u)\alpha}{v_0}}& \mathrm{if}\ v_1 = 0.
\end{cases}
\end{equation}
The population growth rate $\alpha$ is then determined by the implicit equation
\begin{equation}
\label{eq:ls42}
2 \int_0^1 p(u) K_0(\alpha, 1, u) \mathrm{d} u = 1.
\end{equation}

Equations \eqref{eq:ls39} and \eqref{eq:ls42} allow for the direct computation of the steady-state solution. We first solve \eqref{eq:ls42} to determine $\alpha$, which is then substituted into \eqref{eq:ls39} to compute $\hat{f}(s, a)$. Moreover, equation \eqref{eq:ls42} imposes constraints on the parameters $v_0, v_1$, the probability density function $p(s)$, and the steady-state population growth rate $\alpha$.

To facilitate comparison with experimental data, where cell cycle age $a$ is often unobservable, we integrate the steady-state solution over $a$ to obtain the steady-state distribution:
\begin{equation}
	\tilde{f}(s) =\int_0^{+\infty}  \hat{f}(s, a)\mathrm{d}a.
\end{equation}
This gives
\begin{equation}
	\label{eq:s9}
	\tilde{f}(s) = \dfrac{2\alpha}{v(s)}\int_{0}^s p(u) K_0(\alpha, s, u)\mathrm{d}u.
\end{equation} 
Furthermore, \eqref{eq:ls40} implies that $\tilde{f}(1) = (v_1 + v_0)/\alpha$. 

The normalization condition is straightforward to verify using the constraint \eqref{eq:ls42}:
\begin{eqnarray*}
\int_0^1 \tilde{f}(s) \mathrm{d} s &=&\int_0^1 \dfrac{2\alpha}{v(s)}\int_0^s p(u) K_0(\alpha, s, u) \mathrm{d} u\mathrm{d} s\\
&=&2\alpha \int_0^1 p(u)\int_u^1 \dfrac{K_0(\alpha, s, u)}{v(s)} \mathrm{d} s\mathrm{d} u\\
&=&2\alpha \int_0^1 p(u) (-\frac{1}{\alpha}) (K_0(\alpha, 1, u) - 1) \mathrm{d} u\\
&=& -2 \int_0^1 p(u) K_0(\alpha, 1, u) \mathrm{d} u + 2\int_0^1 p(u)\mathrm{d} u \\
&=&1.
\end{eqnarray*}

In summary, we present the following result for the case of linear growth rates. 

\begin{theorem}
\label{th:3.4}
Consider the sizer mechanism described by equation \eqref{eq:size3}. When the cell growth rate is $v(s) = v_0 + v_1 s$, the following conclusions hold:
\begin{enumerate}
\item[\rm{(1)}] The cell size distribution $f(s, a, t)$ is expressed as:
\begin{equation}
\label{eq:s27}
f(s, a, t) = 
\begin{cases}
\displaystyle 2 p(\gamma_0(s, a)) \alpha(t-a)e^{-\int_{t-a}^t (\alpha(t') + v_1) \mathrm{d} t'} & a < t, \gamma_0(s, a)\geq 0\vspace{1mm}\\
\displaystyle g(\gamma_{a-t}(s, a), a-t)e^{-\int_0^t (\alpha(t') + v_1) \mathrm{d} t'} & a \geq t, \gamma_{a-t}(s, a) \geq 0\vspace{1mm}\\
0 & \text{otherwise},
\end{cases}
\end{equation}
where 
$$\gamma_0(s, a) = s e^{-v_1 a} - \frac{v_0}{v_1} (1 - e^{-v_1 a}),$$ 
and 
\begin{equation}
\alpha(t) = \int_{\gamma_0(1, a) \geq 0} v(1) f(1, a, t) \mathrm{d} a.
\end{equation}
\item[\rm{(2)}] The steady-state solution $\hat{f}(s, a)$ is given by
\begin{equation}
\hat{f}(s, a) = 
\begin{cases}
\displaystyle 2 \alpha p(\gamma_0(s, a)) e^{-(\alpha + v_1) a} & s > \frac{v_0}{v_1} (e^{v_1 a} - 1)\vspace{1mm}\\
\displaystyle 0 & \text{otherwise}.
\end{cases}\\
\end{equation}
Define the function $K_0(\alpha, s, u)$ as:
\begin{equation}
K_0(\alpha, s, u) = 
\begin{cases}
\displaystyle \left( \frac{v_0 + s v_1}{v_0 + u v_1}\right)^{-\alpha/v_1},& \mathrm{if}\ v_1 \not= 0\vspace{1mm}\\
\displaystyle e^{-\frac{(s-u)\alpha}{v_0}},& \mathrm{if}\ v_1 = 0.
\end{cases}
\end{equation}
The population growth rate $\alpha$ is determined by the implicit equation:
\begin{equation}
\label{eq:s6}
2\int_0^1 p(u) K_0(\alpha, 1, u) \mathrm{d} u = 1.
\end{equation}
Furthermore, this equation imposes constraints on the parameters $v_0, v_1$, the probability function $p(s)$, and the population growth rate $\alpha$ at the steady state.
\item[\rm{(3)}]
The steady-state size distribution $\tilde{f}(s)$ is given by:
\begin{equation}
\label{eq:sl51}
\tilde{f}(s) = \dfrac{2\alpha}{v(s)}\int_0^s p(u) K_0(\alpha, s, u) \mathrm{d} u.
\end{equation} 
The distribution satisfies the normalization condition 
\begin{equation}
\int_0^1 \tilde{f}(s) \mathrm{d} s = 1.
\end{equation}
\end{enumerate}
\end{theorem}

To verify the theoretical results, we numerically calculate the steady-state distribution based on Theorem \ref{th:3.4} and compare it with stochastic simulations. 

In simulations, we assumed that the sizes of newborn cells follow a beta distribution. Thus, the probability density function $p(s)$ is given by
\begin{equation}
\label{eq:betad}
p(s) = \dfrac{s^{\bar{a} -1} (1-s)^{\bar{b}-1}}{B(\bar{a}, \bar{b})},\quad B(\bar{a}, \bar{b}) = \dfrac{\Gamma(\bar{a}) \Gamma(\bar{b})}{\Gamma(\bar{a} + \bar{b})},
\end{equation}
where $\Gamma(\cdot)$ represents the gamma function. The shape parameters $a'$ and $b'$ define the distribution. Here, we assume symmetric division such that $\bar{a} = \bar{b}$, giving $\mathrm{E}[s] = \frac{1}{2}$ and $\mathrm{var}[s] = \frac{1}{4 (2 \bar{a} +1)}$.  
  
\begin{figure}[htbp]
	\centering	
	\includegraphics[width=14cm]{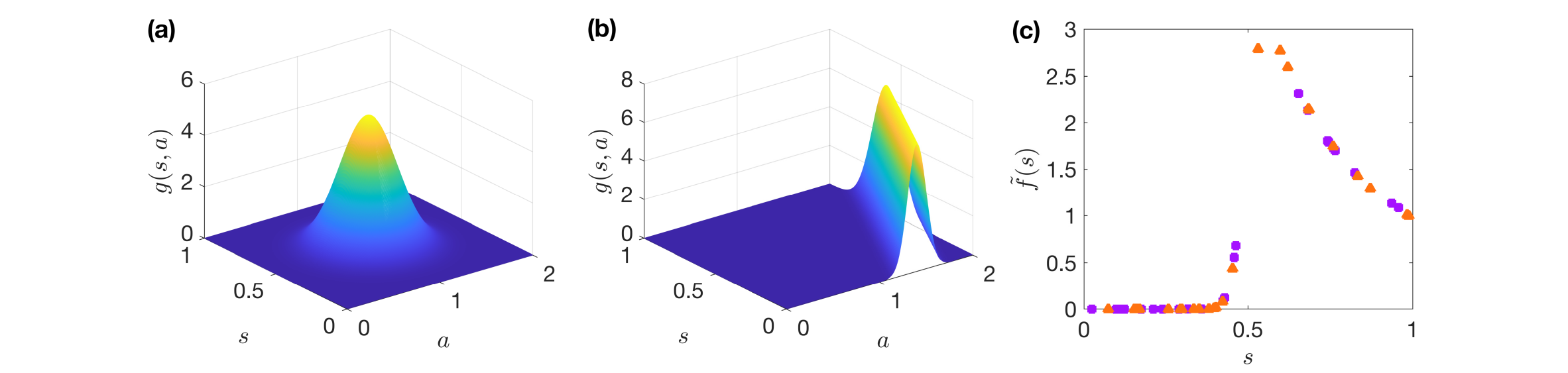}
	\caption{\textbf{Steady-state cell size distribution.} \textbf{(a)} Initial condition $g(s, a)$ defined by \eqref{eq:g1}. \textbf{(b)} Initial condition $g(s, a)$ defined by \eqref{eq:g2}. \textbf{(c)} The steady-state solutions obtained from \eqref{eq:s27} under different initial conditions (marked by distinct markers). In the simulations, $v_0 = 0, v_1 = 2$, and $p(s)$ is taken as beta-distribution \eqref{eq:betad} with $\bar{a} = \bar{b} = 96$.}
	\label{fig:13}
\end{figure}

To examine the solution from different initial conditions, we consider two distinct initial cell size distributions $g(s, a)$:
\begin{equation}
\label{eq:g1}
g(s, a) = \dfrac{1}{2\pi C}e^{-((s - \mu_1)^2/\sigma_1^2 + (a-\mu_2)^2/\sigma_2^2)}, \ \mathrm{where}\ 
\mu_1 = \frac{1}{2}, \mu_2 = 1, \sigma_1 = \frac{1}{8}, \sigma_2 = \frac{1}{4}, 
\end{equation}  
and
\begin{equation}
\label{eq:g2}
g(s, a) = \dfrac{1}{2\pi C}e^{-(s - \mu(a))^2/\sigma^2}, \ \mathrm{where}\ \mu(a) = 2 a - 1.4, \sigma = 0.1.
\end{equation}  
Here, $(s, a) \in D = [0, 1]\times [0, 2]$ and $C$ is a normalized constant so that $\iint_D g(s, a) \mathrm{d} a \mathrm{d}s = 1$. The initial functions $g(s, a)$ are shown at Fig. \ref{fig:13}a-b.

We set the cell growth rate $v(s) = 2 s$ and calculated the solution following the iteration equation \eqref{eq:s27}, using the initial distributions $g(s, a)$ given by \eqref{eq:g1} and \eqref{eq:g2}, respectively. Fig. \ref{fig:13}c displays the steady-state distribution $\tilde{f}(s)$ as $t\to +\infty$ for the two initial conditions. The solutions converge to the same steady state regardless of the initial distribution. 

To examine how changes in $v_0$ and $v_1$ affect the steady-state distribution, we vary the parameters and calculate the size distribution $\tilde{f}(s)$. 

\begin{figure}[htbp]
	\centering
	\includegraphics[width=10cm]{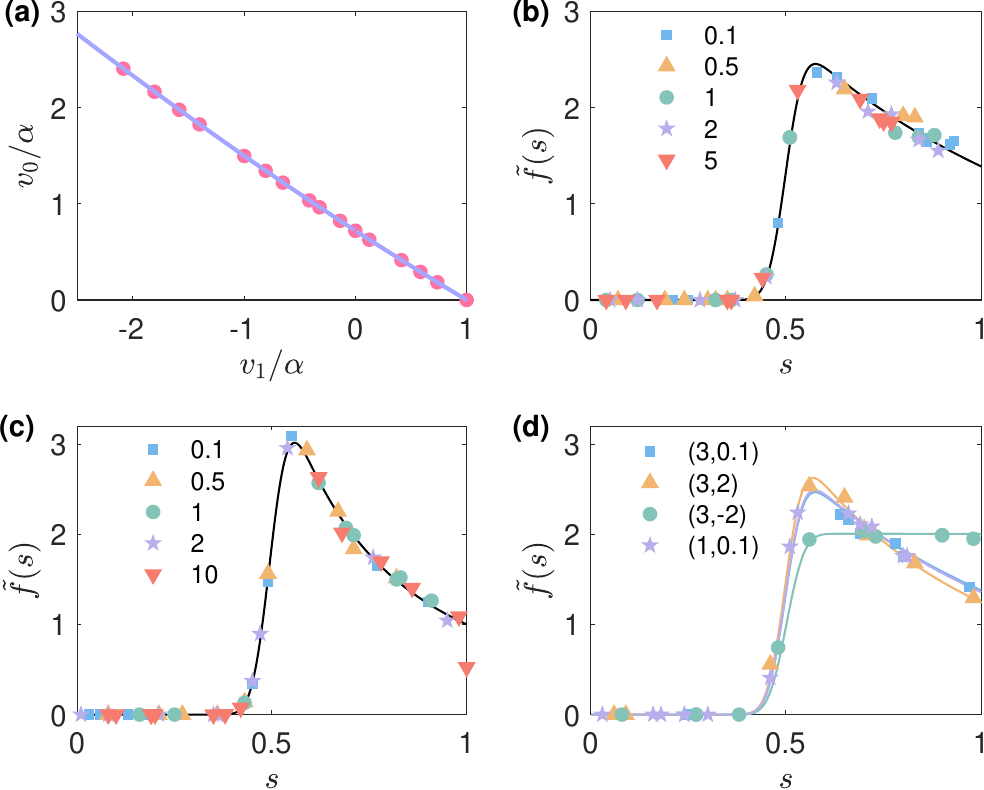}
	\caption{\textbf{Numerical simulation of linear growth rates.} \textbf{(a)} The relationshi between $\frac{v_1}{\alpha}$ and $\frac{v_0}{\alpha}$. \textbf{(b)} Cell size distribution. The solid line represents the theoretical density function \eqref{eq:3.64}. The dots are obtained from cell-based stochastic simulation with growth rate $v(s) = v_0$ and $v_0 = 0.1, 0.5, 1, 2, 5$. \textbf{(c)} Cell size distribution. The solid line represents the theoretical density function \eqref{eq:3.64}. The dots are obtained from cell-based stochastic simulation with growth rate $v(s) = v_1 s$ and $v_1 = 0.1, 0.5, 1, 2, 10$.  \textbf{(d)} Cell size distributions. Different colors represent different growth rates $v(s) = v_0 + v_1 s$, with $(v_0, v_1) = (3, 0.1),  (3, 2), (3, -2)$, and $(1, 0.1)$, respectively. Solid lines are calculated following  \eqref{eq:3.64}, and the dots are obtained from cell-based stochastic simulations. The function $p(u)$ is taken as a beta distribution density function \eqref{eq:betad} with parameters $\bar{a} = \bar{b} = 96$. }
	\label{fig:2}
\end{figure}

First, the constraint given by \eqref{eq:s6} implies
$$
2\int_0^1 p(u) \left(\frac{v_0/\alpha + v_1/\alpha}{v_0/\alpha + u v_1/\alpha}\right)^{-\alpha/v_1}\mathrm{d} u = 1.
$$ 
Using the mean value theorem for definite integrals, there exists $\tilde{u}$ in $[0, 1]$ such that
\begin{equation}
\label{eq:sl55}
2 \left(\frac{v_0/\alpha + v_1/\alpha}{v_0/\alpha + \tilde{u} v_1/\alpha}\right)^{-\alpha/v_1} = 1.
\end{equation}
This equation establishes a constant between $v_0/\alpha$ and $v_1/\alpha$.  The constraint is shown in Fig. \ref{fig:2}a, by assuming $p(s)$ follows the same beta-distribution as in Fig. \ref{fig:13}. Note that $v_1$ can take negative values, provided $v(s) > 0$ for $s\in (0, 1)$, i.e, $v_1 > -v_0$.
 
There are two straightforward results from \eqref{eq:sl55}. When $v_1  = 0$, i.e., the cell size increases linearly with the cell cycle age, \eqref{eq:sl55} simplifies to:
\begin{equation}
\alpha = \beta_0 v_0,\  \mathrm{where}\ \beta_0 = \frac{\log 2}{1 - \tilde{u}}.
\end{equation}
The steady-state distribution given by \eqref{eq:sl51} becomes:
\begin{equation}
\tilde{f}(s) = 2 \beta_0\int_0^s p(u) e^{-\beta_0(s - u)}\mathrm{d} u.
\end{equation}
Interestingly, the steady-state distribution is independent of the growth rate ($v_0)$ of cell sizes. This is clearly illustrated in Fig. \ref{fig:2}b.

When $v_0 = 0$, i.e., the cell size increases exponentially with the cell cycle age, \eqref{eq:sl55} gives
 \begin{equation}
 \alpha = \beta_1 v_1,\ \mathrm{where}\ \beta_1 = \dfrac{\log 2}{\log (1/\tilde{u})}.
 \end{equation}
The steady-state distribution given by \eqref{eq:sl51} becomes:
 \begin{equation}
 \label{eq:3.64}
 \tilde{f}(s) = 2\beta_1 s^{-(1+\beta_1)}\int_0^s p(u) u^{\beta_1}\mathrm{d} u.
 \end{equation}
Similar to the linear growth case, the steady-state distribution is independent of the exponential growth rate ($v_1$). This is shown in Fig. \ref{fig:2}c.
 
When both $v_0$ and $v_1$ are nonzero, the steady-state distribution depends explicitly on the values of $v_0$ and $v_1$, as shown in Fig. \ref{fig:2}d.  

\subsubsection{Logistic cell size growth rate}
When cell sizes follow a logistic growth pattern, the growth rate is given by
\begin{equation}
\label{eq:lograte}
v(s) = v_1 s(1-\frac{s}{v_0}),
\end{equation}
where $v_1 > 0$, and $v_0 > 1$ represents the cell size capacity. Using this growth rate, we derive the following expressions:
\begin{eqnarray*}
u(s, a) &=& \log \dfrac{s}{v_0 - s} - a v_1,\\
w(\varphi_2, a) &=& \dfrac{v_0 e^{\varphi_2 + a v_1}}{1 + e^{\varphi_2 + a v_1}},\\
\gamma_0(s, a) &=& \dfrac{s e^{-a v_1}}{1 - \frac{s}{v_0}(1 - e^{- a v_1})},\\
\gamma_{a-t}(s, a) &=& \dfrac{s e^{-t v_1}}{1 - \frac{s}{v_0}(1 - e^{-t v_1})},\\
\psi(a; \varphi_1, \varphi_2) &=& \alpha (a + \varphi_1) + v_1 \dfrac{1 - e^{\varphi_2 + a v_1}}{1 + e^{\varphi_2 + a v_1}},\\
\psi(a'; t-a, u(s, a)) &=& \alpha (a' + t - a) + v_1 (1 - \frac{2s}{v_0}),\\
\psi(a - t'; t-a, u(s, a)) &=& \alpha(t - t') + v_1 (1 - \frac{2s}{v_0}).
\end{eqnarray*}
Thus, using these expressions, the solution of the governing equation from \eqref{eq:schara} is given by:  
\begin{equation}
f(s, a, t) = 
\begin{cases}
2 p(\gamma_0(s, a)) \alpha(t - a) e^{-\int_{t-a}^t  (\alpha (t') + v_1 (1 - \frac{2s}{v_0})) \mathrm{d} t'}
& a < t,0<s \leq 1 \vspace{1mm}\\
g(\gamma_{a-t}(s, a), a-t) e^{-\int_0^t (\alpha(t') + v_1 (1 - \frac{2s}{v_0}))\mathrm{d} t'} &a \geq t, 0< s\leq 1\vspace{1mm}\\
0&\text{otherwise}.
\end{cases}
\end{equation}
Here, the population growth rate $\alpha(t)$ is defined as:
$$
\alpha(t) = \int_0^{+\infty} v(1) f(1, a, t) \mathrm{d} a.
$$

Let $t\to+\infty$, and consider the steady-state distribution:
\begin{equation}
\label{eq:s61}
\hat{f}(s, a) = 2\alpha e^{-(\alpha + v_1 (1 - \frac{2 s}{v_0}))a}p(\gamma_0(s, a)),\quad (0\leq s \leq 1, a \geq 0)
\end{equation} 
where the population growth rate is given by:
\begin{equation}
\label{eq:s62}
\alpha = v(1) \int_0^{+\infty} \hat{f}(1, a)\mathrm{d} a.
\end{equation}

To facilitate further analysis, we introduce the function $K_1(\alpha, s, u)$:
\begin{equation}
K_1(\alpha, s, u) = \left(\frac{u}{s}\right)^{(\frac{\alpha}{v_1} - \frac{2 s}{v_0})}  \left(\frac{1 - u/v_0}{1 - s/v_0}\right)^{-(\frac{\alpha}{v_1} + 2 (1 - \frac{s}{v_0}))}.
\end{equation}

Substituting \eqref{eq:s61} into \eqref{eq:s62}, we obtain:
$$
\alpha = 2 \alpha v(1)\int_0^{+\infty} e^{-(\alpha + v_1 (1 - \frac{2}{v_0}))a}p(\gamma_0(1, a)) \mathrm{d} a. 
$$
Letting $u = \gamma_0(1, a)$ and applying the function $K_1(\alpha, s, u)$, the population growth rate $\alpha$ satisfies the implicit equation:
\begin{equation}
\label{eq:s63}
1 = 2 \int_0^1 K_1(\alpha, 1, u) p(u) \mathrm{d} u.
\end{equation}

Thus, the growth rate $\alpha$ is determined by solving \eqref{eq:s63}. Once $\alpha$ is known, the steady-state distribution is given by \eqref{eq:s61}. 

Next, we compute the marginal steady-state distribution $\tilde{f}(s)$ by integrating \eqref{eq:s61} over $a$:
$$
\tilde{f}(s) = 2\alpha \int_0^{+\infty} e^{-(\alpha  + v_1 (1 - \frac{2 s}{v_0}))a} p(\gamma_0(s, a)) \mathrm{d} a.
$$
After simplifying, the steady-state distribution becomes:
\begin{equation}
\tilde{f}(s) = \dfrac{2\alpha}{v(s)}\int_0^s K_1(\alpha, s, u) p(u) \mathrm{d} u.
\end{equation}
where $v(s) = v_1 s (1 - s/v_0)$ is the logistic growth rate.

In summary, we present the following theorem for the case where cell sizes follow logistic growth.
\begin{theorem}
Consider the sizer mechanisms described by \eqref{eq:size3}. When the cell growth rate is $v(s)= v_1 (1 - s/v_0)$, where $v_1 > 0$ and $v_0 > 1$, the following conclusions hold:
\begin{enumerate}
\item[\rm{(1)}] The cell size distribution $f(s, a,t )$ is expressed as:
\begin{equation}
f(s, a, t) = 
\begin{cases}
2 p(\gamma_0(s, a)) \alpha(t - a) e^{-\int_0^a \psi(a; t-a, u(s, a)) \mathrm{d} a'}
& a < t,0<s\leq 1 \vspace{1mm}\\
g(\gamma_{a-t}(s, a), a-t) e^{-\int_0^t \psi(a - t'; t - a, u(s, a))\mathrm{d} t'} &a \geq t, 0< s\leq 1\\
0&\text{otherwise}.
\end{cases}
\end{equation}
where
$$
\alpha(t) = \int_0^{+\infty} v(1) f(1, a, t) \mathrm{d} a.
$$
\item[\rm{(2)}] The steady-state solution $\hat{f}(s, a)$ is given by
\begin{equation}
\hat{f}(s, a) = 2\alpha e^{-(\alpha + v_1 (1 - \frac{2 s}{v_0}))a}p(\gamma_0(s, a)),\quad (0\leq s \leq 1, a \geq 0).
\end{equation} 
Define the function $K_1(\alpha, s, u)$ as:
\begin{equation}
K_1(\alpha, s, u) = \left(\frac{u}{s}\right)^{(\frac{\alpha}{v_1} - \frac{2 s}{v_0})}  \left(\frac{1 - u/v_0}{1 - s/v_0}\right)^{-(\frac{\alpha}{v_1} + 2 (1 - \frac{s}{v_0}))}.
\end{equation}
The population growth rate $\alpha$ is determined by the implicit equation:
\begin{equation}
1 = 2 \int_0^1 K_1(\alpha, 1, u) p(u) \mathrm{d} u.
\end{equation}
\item[\rm{(3)}] The steady-state size distribution $\tilde{f}(s)$ is expressed as:
\begin{equation}
\tilde{f}(s) = \dfrac{2\alpha}{s v(s)}\int_0^s K_1(\alpha, s, u) p(u) \mathrm{d} u.
\end{equation}
\end{enumerate} 
\end{theorem}

\begin{figure}[htbp]
	\centering
	\includegraphics[width=12cm]{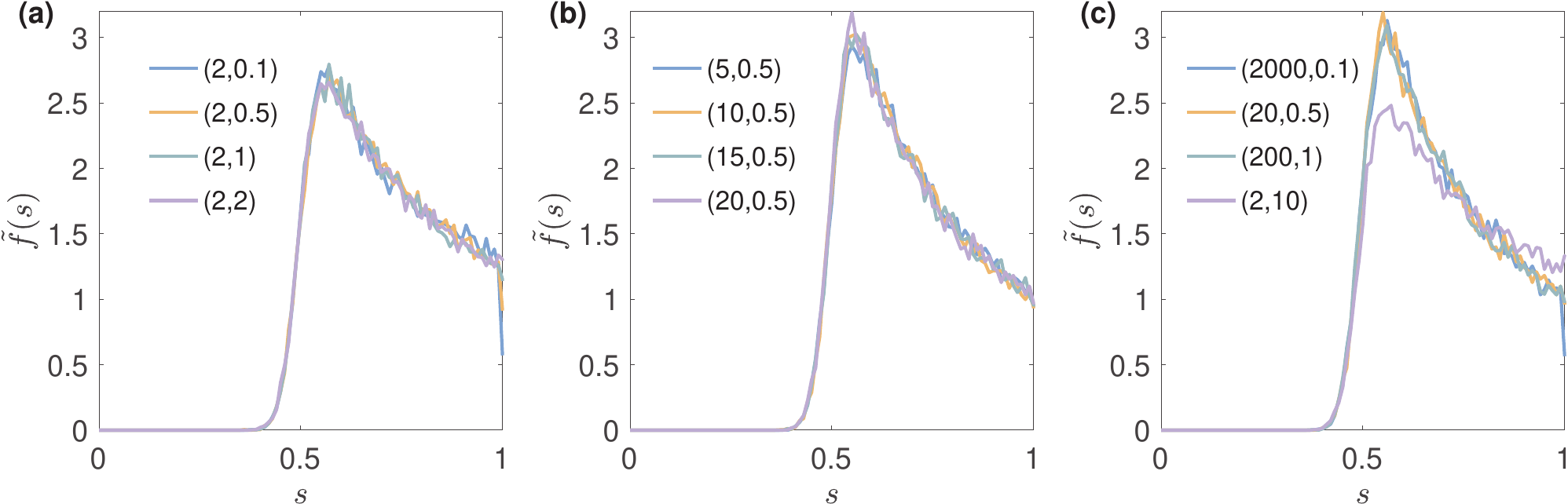}
	\caption{\textbf{Size distribution with logistic cell growth rate \eqref{eq:lograte}.} \textbf{(a)}  Parameters $v_0 = 2$ and $v_1 = 0.1, 0.5, 1, 2$. \textbf{(b)} Parameters $v_1=0.5$ and $v_ 0 = 5,10,15, 20$. \textbf{(c)} Parameters $(v_0, v_1)  = (2000, 0.1), (20, 0.5), (200, 1), (2, 10)$.}
	\label{fig:4}
\end{figure}

Fig. \ref{fig:4} illustrates the steady-state size distributions obtained from stochastic simulations. The results reveal a size distribution shape resembling that of linear growth: the frequency increases from $s =0.4$ to $s = 0.6$, then decreases toward the division size. Additionally, varying $v_1$ (Fig. \ref{fig:4}a) or $v_0$ (Fig. \ref{fig:4}b) alone does not significantly alter the overall shape of the distribution, while changes in both $v_0$ and $v_1$ may affect the size distribution (Fig. \ref{fig:4}c).

\subsubsection{Soft control condition}
In the previous discussions, we assumed a hard control condition, where cell division occurs exclusively when the cell size reaches $s = 1$. Here, we extend this framework to a soft control condition, allowing cell division to occur probabilistically based on the cell size. Specifically, we assume that the division probability follows a normal distribution with a mean of $s=1$ and variance $\sigma^2$. The division probability is expressed as:
\begin{equation}
\label{eq:sizesoft}
\phi(s) = \dfrac{1}{\sqrt{2\pi}\sigma}e^{-\frac{(s-1)^2}{2\sigma^2}}.
\end{equation}

Under this soft control condition, the governing equations become:
\begin{equation}
\left\{
\begin{aligned}
&\dfrac{\partial f(s, a, t)}{\partial t} + \dfrac{\partial f(s, a, t)}{\partial a} + \dfrac{\partial (v(s) f(s, a, t))}{\partial s} = -(\alpha(t) + \phi(s))f(s, a, t),\\
&\alpha(t) = \int_0^{+\infty} f(s, 0, t) \mathrm{d} s - \int_0^{+\infty}\int_0^{+\infty} \phi(s) f(s, a, t) \mathrm{d} a \mathrm{d} s,\\
&f(s, 0, t) = 2 \int_0^{+\infty} \int_0^{+\infty} s' p(s/s')  \phi(s') v(s') f(s', a, t)\mathrm{d}s' \mathrm{d} a,\\
&f(0, a, t) = f(s, +\infty, t) = 0.
\end{aligned}
\right.
\end{equation}
 
\begin{figure}[htbp]
	\centering
	\includegraphics[width=12cm]{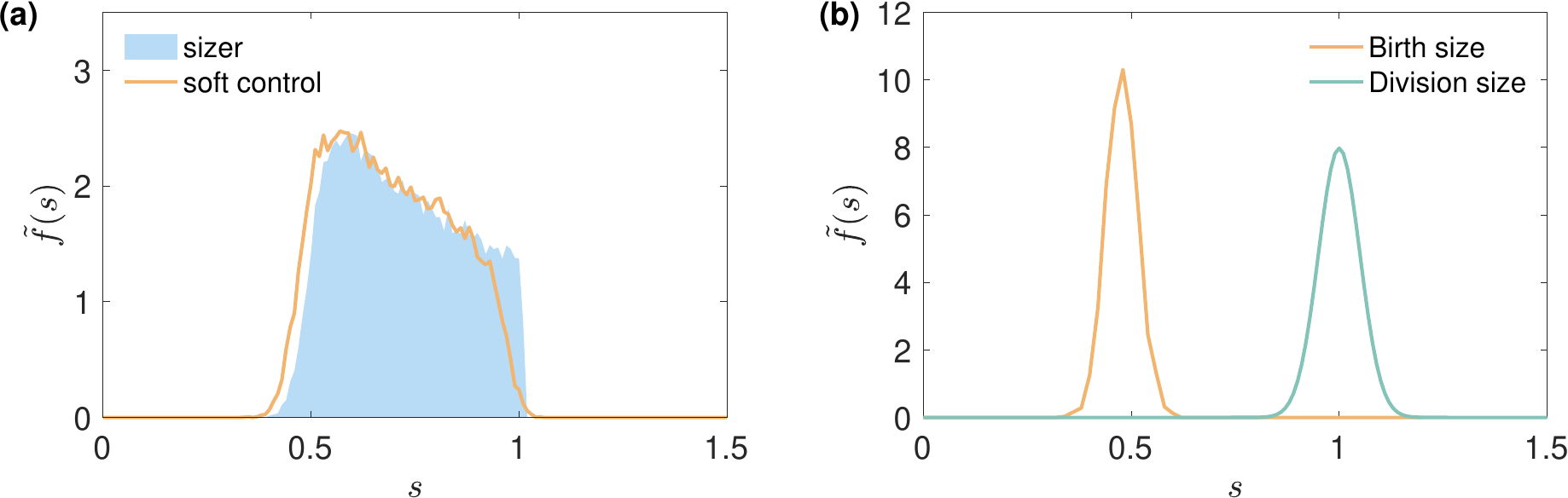}
	\caption{Size distribution. (a) Steady-state cell distributions under hard sizer and soft control conditions. Size distributions were obtained from stochastic simulations with the same growth rate $v = v_0 + v_1 s$ with $v_0 = 1, v_1 = 0.1$. The sort control is taken as \eqref{eq:sizesoft} with $\sigma = 0.02$. The function $p(u)$ is taken as a beta distribution density function \eqref{eq:betad} with parameters $\bar{a} = \bar{b} = 96$. (b) Birth size distribution and division size distribution under soft control conditions.}
	\label{fig:5}
\end{figure}

To explore the effects of the soft control condition, we assumed a linear cell growth model, $v(s) = v_0 + v_1 s$. Using individual-cell-based stochastic simulations, we investigated the steady-state cell size distribution under both the sizer and soft control conditions using the same parameters $v_0$ and $v_1$ (Fig. \ref{fig:5}a). The two controls yield similar steady-state size distributions; however, the soft control condition produces smoother distributions with extended tails near $s=1$, whereas the hard control condition enforces a strict boundary at $0 < s < 1$. Fig. \ref{fig:5}b shows the density distributions of both birth size and division size under the soft control condition.

\subsection{Timer mechanism}

\subsubsection{Exact solution}
For the timer mechanism, cell division occurs strictly at $a = T$, and the governing equation is provided by \eqref{eq:timerequation}. 

To derive the exact solution of \eqref{eq:timerequation}, we follow the methodology and notations used in the sizer mechanism. The solution can be expressed in the form:
\begin{equation}
\label{eq:tf0}
f(s, a, t) = \Psi(t-a, u(s, a))e^{-\int_0^a \psi(a'; t-a, u(s, a))\mathrm{d} a'}.
\end{equation}
To determine the function $\Psi$, we use the boundary condition at $a = 0$:
$$
f(s, 0, t) = \Psi(t, u(s, 0)) = 2 \int_s^{+\infty}p(s, s') f(s', T, t) \mathrm{d} s'.
$$
Substituting $s = w(u(s, 0), 0)$, we obtain
$$
\Psi(t, u) = 2 \int_{w(u, 0)}^{+\infty} p(w(u, 0), s') f(s', T, t) \mathrm{d} s'.
$$

When $t  > a$, the solution \eqref{eq:tf0} can be expressed as:
$$
f(s, a, t) = 2 e^{-\int_0^a \psi(a'; t-a, u(s, a))\mathrm{d} a'} \int_{\gamma_0(s, a)}^{+\infty} p(\gamma_0(s,a), s')f(s', T, t-a) \mathrm{d} s',
$$
where $\gamma_{0}(s, a) = w(u(s, a), 0)$.

When $t < a$, using a similar argument as in the sizer mechanism, the solution \eqref{eq:tf0} becomes
$$
f(s, a, t) = g(\gamma_{a-t}(s, a), a-t) e^{-\int_0^t \psi(a-t'; t-a, u(s, a))\mathrm{d} t'},
$$
where $\gamma_{a-t}(s, a) = w(u(s, a), a-t)$.

We summarize the exact solution in the following theorem: 
\begin{theorem}
\label{th:3.6}
For the timer mechanism, the solution of \eqref{eq:timerequation} is given by:
\begin{equation}
	\label{eq:timer3}
	 f(s, a, t) =
	\begin{cases}
\displaystyle	
\begin{aligned}
&2 e^{-\int_0^a \psi(a'; t-a, u(s, a))\mathrm{d} a'}  \\
& \quad \times \int_{\gamma_0(s, a)}^{+\infty} p(\gamma_0(s,a), s')f(s', T, t-a) \mathrm{d} s' 
\end{aligned}
& a < t, \gamma_0(s, a) \geq 0\vspace{1mm}\\
\displaystyle	g(\gamma_{a-t}(s, a), a-t) e^{-\int_0^t \psi(a-t'; t-a, u(s, a))\mathrm{d} t'}& a \geq t, \gamma_{a-t}(s, a) \geq 0\vspace{1mm}\\
\displaystyle	0 & \text{otherwise}.
	\end{cases}
\end{equation}
The population growth rate is given by:
\begin{equation}
\alpha(t) = \int_0^{+\infty} f(s, T, t)\mathrm{d} s.
\end{equation}
Here, $s \geq 0$, $0\leq a \leq T$, and $t \geq 0$. The functions $\gamma_{a'}(s, a)$ and $\psi(a; \varphi_1, \varphi_2)$ are defined in Theorem \ref{th:3.1}.
\end{theorem}

Usually, similar to the sizer mechanism, we can assume that the probability density function $p(s, s')$ depends only on the ratio $s/s'$ so that $p(s, s') = p(s/s')s'$. Where $p(u)$ gives the probability density of the ratio $u = s/s'$, and satisfies the normalization condition:
\begin{equation}
\int_0^1 p(u) \mathrm{d} u = 1.
\end{equation} 

In Theorem \ref{th:3.6}, the function $\psi$ gives:
$$
\psi(a'; t-a, u(s, a)) = \alpha(a' + t - a) + v_s'(w(s, a), a'), a')
$$
and
$$
\psi(a - t'; t-a, u(s, a)) = \alpha(t-t') + v_s'(w(s, a), a-t'), a-t').
$$
Here, $t -a \leq a' + t - a\leq t$ when $0\leq a' \leq a$, and $0\leq t-t' \leq t$ when $0\leq t'\leq t$. Thus, the solution $f(s, a, t)$ can be calculated iteratively using \eqref{eq:timer3}, starting with the initial conditions and applying the boundary conditions at $a = 0$ and $a = T$.

\subsubsection{Linear growth rates}

When the cell growth rate depends linearly on the cell size $s$, i.e., $v(s) = v_0 + v_1 s$, the change in the cell size with respect to the cell cycle age $a$ is given by
$$
s(a) = s(0) e^{v_1 a} + \frac{v_0}{v_1} (e^{v_1 a} -1).
$$
Let $s_n$ denote the size of a newborn cell at the $n$-th cycle. Assuming that the cell size is halved at division, the iterative relationship is:
\begin{equation}
\label{eq:74}
s_{n+1} = \dfrac{1}{2} e^{v_1 T} s_n  + \dfrac{1}{2}\frac{v_0}{v_1}(e^{v_1 T} - 1),
\end{equation}
where $T$ is the cell cycle duration. 

From \eqref{eq:74}, the cell size stabilizes as $n\to\infty$ only when $e^{v_1 T} < 2$. Under this condition, the stable newborn cell size is:
$$
s^* = \dfrac{v_0}{v_1} \frac{e^{v_1 T}-1}{2 - e^{v_1 T}}.
$$
Thus, linear growth rates result in a stable positive cell size only when $1 < e^{v_1 T} < 2$ and $v_0 > 0$. Specifically, when $v_0 = 0$ (exponential growth), the timer mechanism cannot maintain a stable positive cell size. This result aligns with experimental observations in biological systems \cite{jun2015cell}.

Under the linear growth rate assumption, we have the following expressions:
$$
\begin{aligned}
&\gamma_0(s, a) = s e^{-v_1 a} - \frac{v_0}{v_1} (1  - e^{-v_1 a}),\\
&\gamma_{t-a}(s, a) = s e^{-v_1 t}  - \frac{v_0}{v_1} (1 - e^{-v_1 t}),\\
&\psi(a'; t - a, u(s, a)) = \alpha(a' + t - a) + v_1,\\
&\psi(a-t'; t-a, u(s, a)) = \alpha(t-t') + v_1.
\end{aligned}
$$
Moreover, the integrals in the solution \eqref{eq:timer3} are simplified as:
$$
\int_0^a (\alpha(a' + t - a) + v_1)\mathrm{d} a' = v_1 a + \int_{t-a}^t \alpha(t') \mathrm{d} t'
$$
and
$$
\int_0^t (\alpha(t-t') + v_1)\mathrm{d} t' = v_1 a + \int_0^t \alpha(t') \mathrm{d} t'.
$$

Substituting these into \eqref{eq:timer3}, the solution $f(s, a, t)$ can be expressed as:
\begin{equation}
	\label{eq:t1}
	 f(s, a, t) =
	\begin{cases}
\displaystyle
\begin{aligned}
&2 e^{-(v_1 a + \int_{t-a}^t \alpha(t')\mathrm{d} t')}\\
&\quad \times \int_{\gamma_0(s, a)}^{+\infty} p(\gamma_0(s, a), s')f(s', T, t-a)\mathrm{d} s' 
\end{aligned}
& a < t, \gamma_0(s, a) \geq 0 \vspace{1mm}\\
\displaystyle	e^{ - (v_1 a + \int_0^t\alpha(t') \mathrm{d} t')}g(\gamma_0(s, t), a-t)& a \geq t, \gamma_0(s, t) \geq 0\\
	0 & \text{otherwise}.
	\end{cases}
\end{equation}
Here,
$$
\begin{aligned}
&\alpha(t) = \int_0^{+\infty} f(s, T, t) \mathrm{d} s,\\
&\gamma_0(s, a) = s e^{-v_1 a} - \frac{v_0}{v_1} (1  - e^{-v_1 a}).
\end{aligned}
$$

As $t\to\infty$, the steady-state distribution $\hat{f}(s, a)$ is given by:
\begin{equation}
\label{eq:76}
\hat{f}(s, a) = 2 e^{-(v_1 + \alpha)a}\int_{\gamma_0(s, a)}^{+\infty} p(\gamma_0(s, a), s') \hat{f}(s', T) \mathrm{d}s',
\end{equation}
where the population growth rate $\alpha$ satisfies
$$
\alpha = \int_0^{+\infty} \hat{f}(s, T) \mathrm{d} s.
$$

At $a = T$, the distribution satisfies 
\begin{equation}
\hat{f}(s, T) = 2 e^{-(v_1 + \alpha)T} \int_{\gamma_0(s, T)}^{+\infty} p(\gamma_0(s, T), s') \hat{f}(s', T)\mathrm{d} s'.
\end{equation}
The population growth rate $\alpha$ is calculated as:
\begin{eqnarray*}
\alpha &=& \int_0^{+\infty} 2 e^{-(v_1+\alpha)T} \int_{\gamma_0(s, T)}^{+\infty} p(\gamma_0(s, T), s') \hat{f}(s', T) \mathrm{d}s' \mathrm{d} s\\
&=&2 e^{-(v_1 + \alpha)T}\int_{\gamma_0(0, T)}^{+\infty} \int_u^{+\infty} p(u, s') \hat{f}(s', T) e^{v_1 T} \mathrm{d} s' \mathrm{d} u\\
&=& 2 e^{-\alpha T} \int_{\gamma_0(0, T)}^{+\infty} \int_u^{+\infty} p(u, s') \hat{f}(s', T) \mathrm{d} s' \mathrm{d} u.
\end{eqnarray*}

Thus, we obtain the equations for $\hat{f}(s, T)$ and $\alpha$ as:
\begin{equation}
\label{eq:77}
\left\{
\begin{aligned}
&\hat{f}(s, T) = 2 e^{-(v_1 + \alpha)T} \int_{\gamma_0(s, T)}^{+\infty} p(\gamma_0(s, T), s') \hat{f}(s', T)\mathrm{d} s',\\
&\alpha = 2 e^{-\alpha T} \int_{\gamma_0(0, T)}^{+\infty} \int_u^{+\infty} p(u, s') \hat{f}(s', T) \mathrm{d} s' \mathrm{d} u.
\end{aligned}
\right.
\end{equation}

The steady-state distribution $\hat{f}(s, T)$ and the population growth rate $\alpha$ can be determined by solving \eqref{eq:77}. Once $\hat{f}(s, T)$ is known, the steady-state distribution for all cell sizes can be obtained from \eqref{eq:76}. This approach provides a numerical scheme for calculating the steady-state distribution.

The steady-state size distribution $\tilde{f}(s)$ is computed by integrating over $a$:
\begin{equation}
\tilde{f}(s) = 2\int_0^T\int_{\gamma_0(s, a)}^{+\infty} e^{-(v_1 + \alpha)a} p(\gamma_0(s, a), s') \hat{f}(s', T) \mathrm{d} s' \mathrm{d} a.
\end{equation}
Using the change of variable $u = \gamma_0(s, a)$, we have:
\begin{eqnarray*}
        \tilde{f}(s) &=& 2 \int_{\gamma_0(s, T)}^s \int_{u}^{+\infty} \hat{f}(s', T) e^{-  \frac{\alpha}{v_1}\log\frac{v_0 + s v_1}{v_0 + u v_1}} p(u, s') \frac{1}{v_0 + s v_1}\mathrm{d} s' \mathrm{d} u \\
&=&\dfrac{2}{v(s)}\int_{\gamma_0(s, T)}^s\int_u^{+\infty}\hat{f}(s', T) \left(\frac{v_0 + s v_1}{v_0 + u v_1}\right)^{-\alpha/v_1} p(u, s') \mathrm{d} s' \mathrm{d} u.
\end{eqnarray*}
Thus, we obtain
\begin{equation}
\tilde{f}(s) = \dfrac{2}{v(s)} \int_{\gamma_0(s, T)}^s \int_{u}^{+\infty} \hat{f}(s', T) K_0(\alpha, s, u) p(u, s') \mathrm{d} s' \mathrm{d} u,
\end{equation}
where 
$$
K_0(\alpha, s, u) = \left(\frac{v_0 + s v_1}{v_0 + u v_1}\right)^{-\alpha/v_1}.
$$

In summary, we have the following theorem:
\begin{theorem}
\label{th:3.7}
For the timer mechanism with a linear growth rate is $v(s) = v_0 + v_1 s$, the following hold:
\begin{enumerate}
\item[\rm{(1)}] The cell size distribution $f(s, a, t)$ is expressed as \eqref{eq:t1}.
\item[\rm{(2)}] The steady-state solution $\hat{f}(s, a)$ is given by \eqref{eq:76}, where $\hat{f}(s, T)$ and $\alpha$ satisfy \eqref{eq:77}.
\item[\rm{(3)}] The steady-state size distribution $\tilde{f}(s)$ is expressed as:
\begin{equation}
\tilde{f}(s) = \dfrac{2}{v(s)}\int_{\gamma_0(s, T)}^s \int_u^{+\infty}\hat{f}(s', T) K_0(\alpha, s, u) p(u, s') \mathrm{d} s' \mathrm{d} u.
\end{equation}
\end{enumerate}
\end{theorem}

To examine the evolution of cell size dynamics, we performed individual-cell-based stochastic simulations starting with $10,000$ cells, each assigned a random cell cycle age $a$ ($0 < a < T$) and an initial cell size $s$ drawn form a normal distribution with mean $\mu  = 0.4 + 0.5 a$ and variance $\sigma = 0.05$. In the simulation, we set $T = 2$ and used a linear cell growth rate $v = v_0 + v_1 s$, with $v_0 = 0.5$ and $v_1 = 0.2$, ensuring $1 < e^{v_1 T} < 2$.  

\begin{figure}[htbp]
	\centering
	\includegraphics[width=11cm]{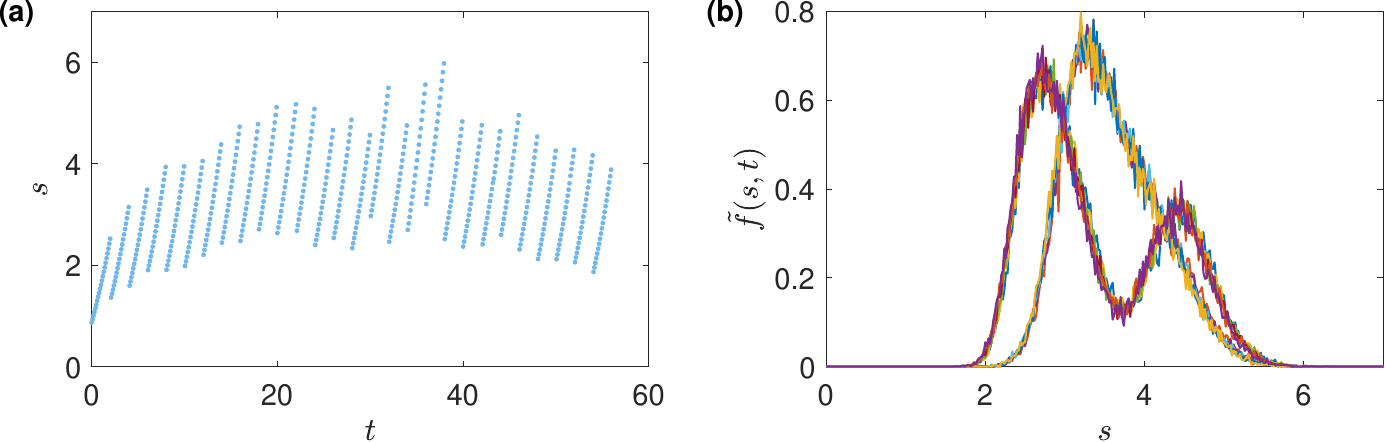}
	\caption{Cell lineage data and size distribution. (a) Evolution of cell size following a single cell. The cell size growth follows a linear growth rate $v = v_0 + v_1 s$ with $v_0 = 0.5 $ and $v_1 = 0.2$. (b) Cell size distributions from $t = 1000$ to $t = 1010$ with a time step $\Delta t = 1$. The function $p(u)$ is taken as a beta distribution density function \eqref{eq:betad} with parameters $\bar{a} = \bar{b} = 96$.}
	\label{fig:28}
\end{figure}

Fig. \ref{fig:28}a shows the size trajectory of a single cell over time, where size increases exponentially during each cycle, and cell division is indicated by a sharp drop in size. Fig. \ref{fig:28}b presents the ell size distribution $\tilde{f}(s, t) = \int_0^T f(s, a, t) \mathrm{d} a$ at time points from $t = 1000$ to $t = 1010$. The figure reveals two distinct distribution patterns within this time window, indicating periodic oscillations in the cell size distribution $\tilde{f}(s, t)$.    

\subsubsection{Soft control condition}

When the control condition is soft so that the division probability follows a normal distribution with a mean $a =T$ and variance $\sigma^2$, the division probability is expressed as
\begin{equation}
\label{eq:timersoft}
\phi(a) = \dfrac{1}{\sqrt{2\pi}\sigma} e^{-\frac{(s-T)^2}{2\sigma^2}}.
\end{equation}
Under this soft control condition, the governing equations become:
\begin{equation}
\left\{
\begin{aligned}
&\dfrac{\partial f(s, a, t)}{\partial t} + \dfrac{\partial f(s, a, t)}{\partial a} + \dfrac{\partial (v(s) f(s, a,t))}{\partial s} = -(\alpha(t) + \phi(a)) f(s, a, t),\\
&\alpha(t) = \int_0^{+\infty} f(s, 0, t) \mathrm{d} s - \int_0^{+\infty}  \int_0^{+\infty} (\phi(a) f(s, a, t)) \mathrm{d} s \mathrm{d} a,\\
&f(s, 0, t) = 2 \int_0^{+\infty} \int_0^{+\infty} p(s, s')\phi(a) f(s', a, t) \mathrm{d} s' \mathrm{d} a,\\
&f(0, a, t) = f(+\infty, a, t) = 0.
\end{aligned}
\right.
\end{equation}

\begin{figure}[htbp]
	\centering
	\includegraphics[width=10cm]{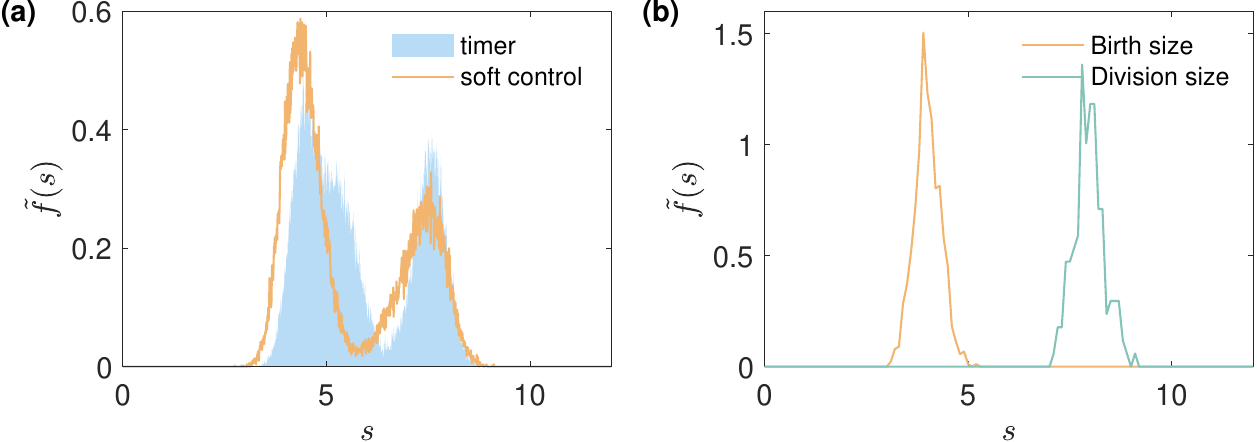}
	\caption{Size distribution. (a) Cell size distributions under hard timer and soft control conditions. Size distributions at the same time point ($t = 1000$) were obtained from stochastic simulations with the same growth rate $v = 0.2$. The sort control is taken as \eqref{eq:timersoft} with $T = 2$ and $\sigma = 0.02$. The function $p(u)$ is taken as a beta distribution density function \eqref{eq:betad} with parameters $\bar{a} = \bar{b} = 96$. (b) Birth size distribution and division size distribution under soft control conditions.}
	\label{fig:7}
\end{figure}

To explore the effects of the soft control condition, we assumed a constant cell growth model, $v(s) = 0.2$. Using individual-cell-based stochastic simulations, we investigated the cell size distribution under both the timer and soft control conditions (Fig. \ref{fig:7}a). The two controls yield similar size distributions; however, the soft control condition yields smaller sizes than the standard timer control. Fig. \ref{fig:7}b shows the density distributions of both birth size and division size under the soft control condition.

\subsection{Adder mechanism} 

\subsubsection{Exact solution}
To derive the exact solution of the adder mechanism, we begin by rewriting the governing equation \eqref{eq:s21} as:
\begin{equation}
\label{eq:83}
\dfrac{\partial f}{\partial t} + v(s, \varsigma) \dfrac{\partial f}{\partial s} + v(s, \varsigma) \dfrac{\partial f}{\partial \varsigma} = -(\alpha(t) + v_\varsigma'(s, \varsigma) + v_s'(s, \varsigma)) f(s, \varsigma, t).
\end{equation}
The corresponding characteristic equation is:
$$
\dfrac{\mathrm{d} t}{1} = \dfrac{\mathrm{d} s}{v(s, \varsigma)} = \dfrac{\mathrm{d} \varsigma}{v(s, \varsigma)} = \dfrac{\mathrm{d} f}{-(\alpha(t) + v_\varsigma'(s, \varsigma) + v_s'(s, \varsigma))f}.
$$

First, it is easy to have a first integral
$$
\varphi_1 = \varsigma - s.
$$
Let $s = \varsigma - \varphi_1$, so that we have:
$$
\dfrac{\mathrm{d} \varsigma}{\mathrm{d} t} = v(\varsigma - \varphi_1, \varsigma),
$$
which leads to the second first integral:
$$
\varphi_2 = u(\varsigma, \varsigma - s, t) = \int_0^\varsigma\dfrac{1}{v(\varsigma' - (\varsigma - s), \varsigma')}\mathrm{d} \varsigma'  - t.
$$

Using the two first integrals:
\begin{equation}
\left\{
\begin{aligned}
&\varphi_1 = \varsigma - s\\
&\varphi_2 = u(\varsigma, \varsigma - s, t), 
\end{aligned}
\right.
\end{equation}
we solve for $t$ and $s$ in terms of $\varphi_1$ and $\varphi_2$:
$$
s = \varsigma - \varphi_1, t = w(\varphi_1, \varphi_2, \varsigma).
$$

Next, we use the differential equation for $f$:
\begin{eqnarray*}
\dfrac{\partial f}{\partial \varsigma} &=& - \dfrac{\alpha(w(\varphi_1, \varphi_2, \varsigma)) + v_\varsigma'(\varsigma - \varphi_1, \varsigma) + v_s'(\varsigma - \varphi_1, \varsigma)}{v(\varsigma - \varphi_1, \varsigma)}f\\
&=& - \psi(\varsigma; \varphi_1, \varphi_2) f,
\end{eqnarray*}
which leads to the third first integral:
$$
\varphi_3 = f e^{-\int_0^\varsigma \psi(\varsigma'; \varphi_1, \varphi_2) \mathrm{d} \varsigma'}. 
$$

Thus, the solution $f(s, \varsigma, t)$ can be written as:
\begin{equation}
\label{eq:85}
\begin{aligned}
f(s, \varsigma, t) &=  \Psi(\varphi_1, \varphi_2) e^{-\int_0^\varsigma \psi(\varsigma'; \varphi_1, \varphi_2)\mathrm{d} \varsigma'}\\
&= \Psi(\varsigma - s, u(\varsigma, \varsigma - s, t)) e^{-\int_0^\varsigma \psi(\varsigma'; \varsigma - s, u(\varsigma, \varsigma - s, t))\mathrm{d} \varsigma'}.
\end{aligned}
\end{equation}
Using the boundary condition at $\varsigma = 0$:
$$
f(s, 0, t) = \Psi(-s, u(0, -s, t)) = 2 \int_{\Delta s}^{+\infty} p(s, s') \varphi(s', \Delta s, t) f(s', \Delta s, t) \mathrm{d} s',
$$
and noting that $t = w(-s, u(0, -s, t), 0)$, we obtain:
$$
\Psi(\varphi_1, \varphi_2) = 2 \int_{\Delta s}^{+\infty} p(-\varphi_1, s') \varphi(s', \Delta s, w(\varphi_1, \varphi_2, 0)) f(s', \Delta s, w(\varphi_1, \varphi_2, 0))\mathrm{d} s'.
$$

Now, define
$$
\tau(s, \varsigma, t) = w(\varsigma -s, u(\varsigma, \varsigma - s, t), 0).
$$
If $\tau(s, \varsigma, t) > 0$, the solution \eqref{eq:85} becomes:
\begin{equation}
\label{eq:86}
\begin{aligned}
f(s, \varsigma, t) = &2 e^{-\int_0^\varsigma \psi(\varsigma'; \varsigma - s, u(\varsigma, \varsigma - s, t))\mathrm{d} \varsigma'} \\
&\quad \times \int_{\Delta s}^{+\infty} p(s - \varsigma, s') \varphi(s', \Delta s, \tau(s, \varsigma, t)) f(s', \Delta s, \tau(s, \varsigma, t))\mathrm{d} s'.
\end{aligned}
\end{equation}

When $\tau(s, \varsigma, t) < 0$, we solve for $\varsigma$ from the equation
\begin{equation}
\label{eq:87}
t = w(\varphi_1, \varphi_2, \varsigma),
\end{equation}
which gives: 
\begin{equation}
\label{eq:88}
\varsigma = \varpi(\varphi_1, \varphi_2, t).
\end{equation}

From the initial condition, we have:
$$
g(s, \varsigma)  = \Psi(\varsigma - s, u(\varsigma, \varsigma - s, 0))e^{-\int_0^\varsigma \psi(\varsigma'; \varsigma - s, u(\varsigma, \varsigma - s, 0)) \mathrm{d} \varsigma'}, 
$$
which gives
$$
\begin{aligned}
\Psi(\varsigma - s, u(\varsigma, \varsigma - s, 0))) &= g(s, \varsigma) e^{\int_0^\varsigma \psi(\varsigma'; \varsigma - s, u(\varsigma, \varsigma - s, 0)) \mathrm{d} \varsigma'}\\
& = g(\varpi(\varsigma - s, u(\varsigma, \varsigma - s, 0), 0) - (\varsigma - s), \varpi(\varsigma - s, u(\varsigma, \varsigma - s, 0), 0))\\
&{}\qquad \times e^{\int_0^{\varpi(\varsigma - s, u(\varsigma, \varsigma - s, 0), 0)} \psi(\varsigma'; \varsigma - s, u(\varsigma, \varsigma - s, 0)) \mathrm{d} \varsigma'}.
\end{aligned}
$$
Thus, we obtain
$$
\Psi(\varphi_1, \varphi_2) = g(\varpi(\varphi_1, \varphi_2, 0) - \varphi_1, \varpi(\varphi_1, \varphi_2, 0))e^{\int_0^{\varpi(\varphi_1, \varphi_2, 0)} \psi(\varsigma'; \varphi_1, \varphi_2) \mathrm{d} \varsigma'}.
$$
Therefore, let
$$
\bar{\varsigma}(s, \varsigma, t) = \varpi(\varsigma - s, u(\varsigma, \varsigma - s, t), 0),
$$
we obtain the following express for the solution:
\begin{equation}
\label{eq:89}
f(s, \varsigma, t) = g(\bar{\varsigma}(s, \varsigma, t) - (\varsigma - s)), \bar{\varsigma}(s, \varsigma, t)) e^{-\int_{\bar{\varsigma}(s, \varsigma, t)}^\varsigma \psi(\varsigma'; \varsigma - s, u(\varsigma, \varsigma - s, t))\mathrm{d} \varsigma'}.
\end{equation}

The initial function $g(s, \varsigma)$ represents the cell number density for cells with size $s$ and size added $\varsigma$ since birth. It is clear that $g(s, \varsigma) > 0$ only when $s \geq \varsigma$. The solution \eqref{eq:89} is valid when $0 < \bar{\varsigma}(s, \varsigma, t) < \varsigma$. From equations \eqref{eq:87}-\eqref{eq:88}, $\bar{\varsigma}(s, \varsigma, t) < \varsigma$ is equivalent to $t = 0$, and $ \bar{\varsigma}(s, \varsigma, t) = 0$ corresponds to the time when $\tau(s, \varsigma, t) < 0$, which is the cell birth time, where the size added $\varsigma = 0$.

Thus, the exact solution can be written as:
\begin{equation}
f(s, \varsigma, t) = \begin{cases}
\displaystyle 
\begin{aligned}
&2 e^{-\int_0^\varsigma \psi(\varsigma'; \varsigma - s, u(\varsigma, \varsigma - s, t))\mathrm{d} \varsigma'} \\
&\times \int_{\Delta s}^{+\infty} p(s - \varsigma, s') \varphi(s', \Delta s, \tau(s, \varsigma, t)) f(s', \Delta s, \tau(s, \varsigma, t))\mathrm{d} s'
\end{aligned}
& 
\begin{aligned}
&s\geq \varsigma\\
&\tau(s, \varsigma, t) \geq 0
\end{aligned} \vspace{1mm}\\
\displaystyle g(\bar{\varsigma}(s, \varsigma, t) - (\varsigma - s)), \bar{\varsigma}(s, \varsigma, t)) e^{-\int_{\bar{\varsigma}(s, \varsigma, t)}^\varsigma \psi(\varsigma'; \varsigma - s, u(\varsigma, \varsigma - s, t))\mathrm{d} \varsigma'} &
\begin{aligned}
&s \geq \varsigma\\
&\tau(s, \varsigma, t)  < 0\\
& 0 < \bar{\varsigma}(s, \varsigma, t) < \varsigma
\end{aligned} \vspace{1mm}\\
0 & \text{otherwise}.
\end{cases}
\end{equation}

\subsubsection{Stochastic simulations}

We applied individual-cell-based stochastic simulations to examine cell size distribution under the adder control mechanism with various cell growth rates.  

Assuming a linear growth rate $v(s, a) = v_0 + v_1 s$, the cell size $s(a)$ at age $a$, with initial value $s(0) = s_b$, is given by 
\begin{equation}
s(a) = \frac{1}{v_1} (v_1 s_b e^{v_1 t} + v_0(e^{v_1 t} - 1)).
\end{equation}
Therefore, a fixed size increment $\Delta s$ corresponds to a division time 
$$T = \frac{1}{v_1} \log \frac{v_0 + v_1 \Delta s}{v_0 + v_1 s_b},$$ 
which depends on the birth size $s_b$.

\begin{figure}[htbp]
	\centering
	\includegraphics[width=12cm]{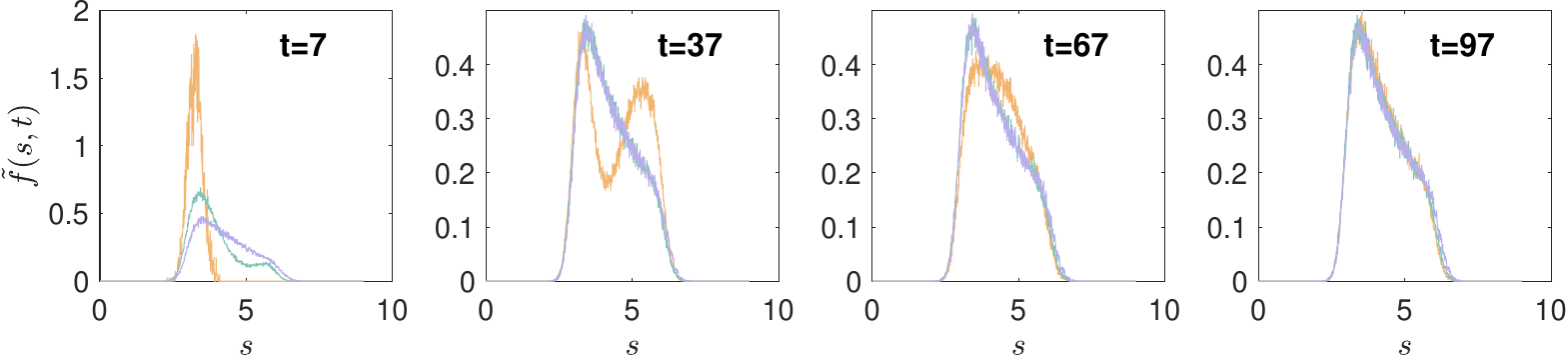}
	\caption{Stochastic simulation of the adder mechanism. The cell growth rate is $v = v_0 + v_1 s$, with $v_0 = 0.5$ and $v_1 = 0.5, 1$, and $2$, respectively. The division probability density $p(s, s')$ is defined as $p(s, s') = p(s/s') s'$, where $p(u)$ the beta distribution density given by \eqref{eq:betad} where $\bar{a} = \bar{b} = 96$. The cell size increment is $\Delta s = 3$.}
	\label{fig:8}
\end{figure}

Fig. \ref{fig:8} shows the cell size distributions at different time points under varying growth rates. While differences are evident at early time points ($t = 7$ and $t = 37$), the distributions eventually converge to a common steady-state profile. This convergence indicates the existence of a stable steady-state distribution under the adder mechanism, independent of the cell growth rate and initial distribution.
 
To validate our model, we applied it to fit experimental cell size distributions of \textit{Escherichia coli} (\textit{E. coli}) reported by Tanouchi et al. \cite{tanouchi2015noity}. In their study, cell length measurements from 279 cell lineages across 70 generations were collected at three different temperatures ($25\circ$C, $27\circ$C, and $37\circ$C) using a mother machine microfluidic device. They proposed a noisy linear map model incorporating negative feedback, in which larger cells tend to divide earlier and smaller cells later, capturing essential features of size control \cite{tanouchi2015noity}. 

\begin{figure}[htbp]
	\centering
	\includegraphics[width=12cm]{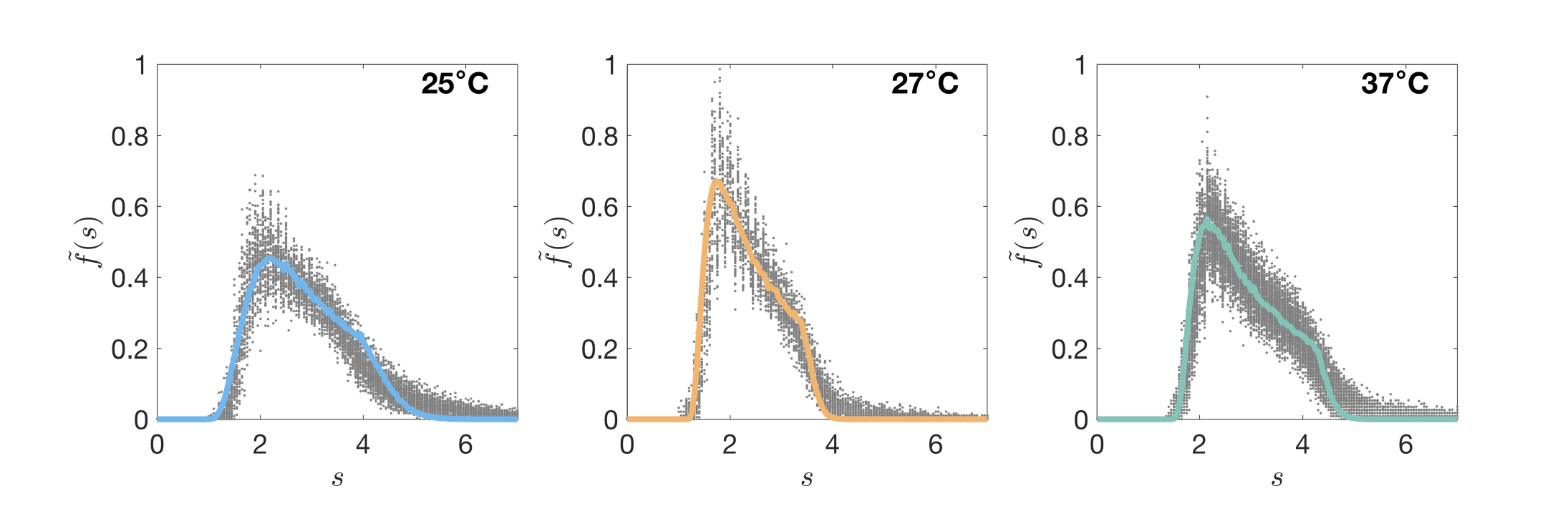}
	\caption{Fitting the experimental cell size distributions. Points represent experimental data at $25^\circ$C, $27^\circ$C, and $37^\circ$C. Solid lines indicate the model simulation results using parameter listed in \eqref{tab:1}.}
	\label{fig:9}
\end{figure}

Fig. \ref{fig:9} shows that the experimental cell size distributions at different temperatures exhibit patterns similar to the steady-state distributions predicted by our simulations in Fig. \ref{fig:8}. Since our results in Fig. \ref{fig:8} suggest that the steady-state distribution is largely insensitive to the cell growth rate, we fixed the growth rate as $v = 0.5 + 0.5 s$ and adjusted the division probability function $p(s, s')$ to match the experimental data.

In real cells, daughter cell sizes are bounded within a range $r s' < s < s'$. We therefore defined the conditional distribution:
$$
p(s, s') = \begin{cases}
(1 - r) s' \times p\left(\frac{s - r s'}{(1 - r) s'}\right), \quad & s_{\min} \leq r s'< s < s'\\
0 \quad & \mathrm{otherwise}
\end{cases}
$$
where $p(u)$ is the beta distribution density as given in \eqref{eq:betad}. We adjusted the parameters $c_0, c_1$, and $\Delta s$ to fit the data at each temperature. 

Fig. \ref{fig:9} shows excellent agreement between the simulation results and the experimental data. The fitted parameter values for each temperature are listed in \eqref{tab:1}. These results indicate that the primary temperature-dependent factor affecting the size distribution under the adder control is the size increment $\Delta s$.
 
\begin{table}[htp]
\caption{Parameter values used to fit the experimental data.}
\begin{center}
\begin{tabular}{cccc}
\hline
Parameter & $25^\circ$C & $27^\circ$C & $37^\circ$C \\
\hline
$r$ & $0.25$ &  $0.35$ & $0.35$ \\ 
$\Delta s$ & $2.7$ & $2.1$ & $2.6$ \\
$\bar{a}$ & $3$ & $3$ & $3$ \\
$\bar{b}$ & $12$ & $25$ & $25$ \\
\hline
\end{tabular}
\end{center}
\label{tab:1}
\end{table}%

\section{Discussions}

In this study, we developed a unified mathematical framework to analyze cell size homeostasis under three fundamental control mechanisms: the sizer, timer, and adder. Our approach is grounded in a first-order partial differential equation model that incorporates the dynamics of cell growth and division through mechanistically distinct boundary conditions. This framework allows us to systematically explore how different size control strategies influence the temporal evolution and steady-state distribution of cell sizes in a proliferating population. 

A major contribution of this work lies in the derivation of exact analytical solutions for each control mechanism using the method of characteristics. These solutions provide valuable insights into how the cell size distribution evolves over time and how it depends on key model parameters, such as the growth rate function, division rules, and the distribution of daughter cell sizes. In particular, we rigorously established the existence and stability of steady-state solutions for the sizer mechanism under biologically realistic assumptions. This theoretical foundation supports a deeper understanding of how size control strategies maintain homeostasis despite intrinsic variability. 

For the size mechanism, we showed that the size distribution at steady state is largely determined by the cell size inheritance probability density function $p(s)$, and is independent of the detailed from the cell growth rate $v(s, a)$. This suggests that division rules play a dominant role in shaping the size distribution in sizer-controlled populations. We also found that nonlinear growth rates, such as exponential or logistic forms, can lead to qualitatively similar steady-state distributions, further emphasizing the robustness of the sizer strategy.

For the timer mechanism, we observed through both analysis and simulation that constant growth rates $v(s, a) = v_0$ can lead to periodic oscillations in the population size distribution. This is due to the synchronized division times imposed by the fixed cell cycle length $T$. When the growth rate is linear in size, i.e., $v(s) = v_0 + v_1 s$, the system may become unstable unless certain conditions are met, indicating potential limitations of pure timer-based control in achieving size homeostasis.

The adder mechanism,  implemented using a two-variable PDE model in terms of cell size $s$ and size added since birth $\varsigma$, was shown to naturally generalize the timer mechanism when the growth rate depends only on cell cycle age. We derived the exact solution of the adder model under general growth functions and examined how the added size influences the cell cycle length and division dynamics. Our simulations confirm that, under varying growth rates, the adder mechanism leads to convergence toward a stable size distribution. 

In addition to exact solutions, we performed extensive individual-based stochastic simulations to validate the theoretical predictions and explore model behavior in biologically relevant scenarios. The simulation results agree well with analytical findings, and also highlight features such as the smoother distribution under soft control conditions, and long-tailed distributions near critical size thresholds, which are difficult to capture analytically. 

From a modeling perspective, this study illustrates how different assumptions about cell growth and division translate into structurally different PDE models, each with distinct mathematical properties. The sizer model has a sharp boundary defined by the division size threshold, while the adder model introduces an added-size variable to track memory from birth. Despite these differences, all three models share a common analytical structure, allowing for unified treatment and comparison.

Several important directions for future work arise from this study. First, our current model treats daughter cells as statistically independent, while in reality, correlations between sibling cells (e.g., equal partitioning of the mother's size) may significantly affect the outcome. Incorporating such correlations would require extending the boundary conditions to account for joint distributions of daughter sizes. Second, although we have considered a fixed added size or cycle length for adder and timer mechanisms, introducing stochasticity in division conditions (e.g., variability in $\Delta s$ or $T$) could better reflect biological reality and deserves deeper mathematical investigation. Lastly, coupling size dynamics with gene regulatory feedback or resource allocation models may uncover deeper principles of growth and division coordination in more complex settings.

In conclusion, this work provides a comprehensive mathematical and computational framework for studying cell size control mechanisms. By integrating analytical theory with stochastic simulations, we offer a set of tools and insights that can guide both experimental design and the development of more sophisticated models in future studies of cellular homeostasis. 
 
\section*{Acknowledegement}
The authors thank Prof. Zhengliang Gao for introducing them to the problem of cell size control, and Ms. Yuhong Zhang for her help in programming.

\bibliographystyle{siamplain}
\bibliography{references}

\begin{thebibliography}{10}

\bibitem{Amodeo:2016aa}
{\sc A.~A. Amodeo and J.~M. Skotheim}, {\em {Cell-size control.}}, Cold Spring
  Harb Perspect Biol, 8 (2016), p.~a019083,
  \url{https://doi.org/10.1101/cshperspect.a019083}.

\bibitem{barber2020cell}
{\sc F.~Barber, A.~Amir, and A.~W. Murray}, {\em {Cell-size regulation in
  budding yeast does not depend on linear accumulation of Whi5}}, Proc Natl
  Acad Sci USA, 117 (2020), pp.~14243--14250,
  \url{https://doi.org/10.1073/pnas.2001255117}.

\bibitem{Bell1968Cell}
{\sc G.~I. Bell}, {\em {Cell growth and division. 3. Conditions for balanced
  exponential growth in a mathematical model}}, Biophys J, 8 (1968),
  pp.~431--444, \url{https://doi.org/10.1016/S0006-3495(68)86498-7}.

\bibitem{Bell1967Cell}
{\sc G.~I. Bell and E.~C. Anderson}, {\em Cell growth and division}, Biophys J,
  7 (1967), pp.~329--351, \url{https://doi.org/10.1016/S0006-3495(67)86592-5}.

\bibitem{Bernard2020Asynchronous}
{\sc T.~Bernard and P.~Gabriel}, {\em Asynchronous exponential growth of the
  growth-fragmentation equation with unbounded fragmentation rate}, J Evol Equ,
  20 (2020), \url{https://doi.org/10.1007/s00028-019-00526-4}.

\bibitem{Cadart:2018aa}
{\sc C.~Cadart, S.~Monnier, J.~Grilli, P.~J. S{\'a}ez, N.~Srivastava, R.~Attia,
  E.~Terriac, B.~Baum, M.~Cosentino-Lagomarsino, and M.~Piel}, {\em {Size
  control in mammalian cells involves modulation of both growth rate and cell
  cycle duration.}}, Nat Commun, 9 (2018), p.~3275,
  \url{https://doi.org/10.1038/s41467-018-05393-0}.

\bibitem{campos2014constant}
{\sc M.~Campos, I.~V. Surovtsev, S.~Kato, A.~Paintdakhi, B.~Beltran, S.~E.
  Ebmeier, and C.~Jacobs-Wagner}, {\em A constant size extension drives
  bacterial cell size homeostasis}, Cell, 159 (2014), pp.~1433--1446,
  \url{https://doi.org/10.1016/j.cell.2014.11.022}.

\bibitem{Chadha:2024aa}
{\sc Y.~Chadha, A.~Khurana, and K.~M. Schmoller}, {\em {Eukaryotic cell size
  regulation and its implications for cellular function and dysfunction.}},
  Physiol Rev, 104 (2024), pp.~1679--1717,
  \url{https://doi.org/10.1152/physrev.00046.2023}.

\bibitem{chandler2017adder}
{\sc D.~Chandler-Brown, K.~M. Schmoller, Y.~Winetraub, and J.~M. Skotheim},
  {\em The adder phenomenon emerges from independent control of pre-and
  post-start phases of the budding yeast cell cycle}, Curr Biol, 27 (2017),
  pp.~2774--2783, \url{https://doi.org/10.1016/j.cub.2017.08.015}.

\bibitem{1962Rate}
{\sc J.~F. Collins and M.~H. Richmond}, {\em Rate of growth of bacillus cereus
  between divisions}, J Gen Microbiol, 28 (1962), p.~15,
  \url{https://doi.org/10.1099/00221287-28-1-15}.

\bibitem{Diekmann1984On}
{\sc O.~Diekmann, H.~J. A.~M. Heijmans, and H.~R. Thieme}, {\em On the
  stability of the cell size distribution}, J Math Biology, 19 (1984),
  pp.~227--248, \url{https://doi.org/10.1007/BF00277748}.

\bibitem{DONACHIE1968Relationship}
{\sc DONACHIE and D.~W.}, {\em {Relationship between Cell Size and Time of
  Initiation of DNA Replication}}, Nature, 219 (1968), pp.~1077--1079,
  \url{https://doi.org/10.1038/2191077a0}.

\bibitem{donnan1983cell}
{\sc L.~Donnan and P.~C. John}, {\em {Cell cycle control by timer and sizer in
  Chlamydomonas}}, Nature, 304 (1983), pp.~630--633,
  \url{https://doi.org/https://doi.org/10.1038/304630a0}.

\bibitem{Facchetti:2017aa}
{\sc G.~Facchetti, F.~Chang, and M.~Howard}, {\em {Controlling cell size
  through sizer mechanisms.}}, Curr Opin Syst Biol, 5 (2017), pp.~86--92,
  \url{https://doi.org/10.1016/j.coisb.2017.08.010}.

\bibitem{Gabriel2019Steady}
{\sc P.~Gabriel and H.~Martin}, {\em Steady distribution of the incremental
  model for bacteria proliferation}, Netw Heterog Media,  (2019),
  \url{https://doi.org/10.3934/NHM.2019008}.

\bibitem{ginzberg2015cell}
{\sc M.~B. Ginzberg, R.~Kafri, and M.~Kirschner}, {\em {On being the right
  (cell) size.}}, Science, 348 (2015), p.~1245075,
  \url{https://doi.org/https://doi.org/10.1126/science.1245075}.

\bibitem{gong2022measuring}
{\sc A.~Gong and M.~Min}, {\em Measuring the size and growth of single cells},
  Biophys Rep, 8 (2022), pp.~150--157,
  \url{https://doi.org/10.52601/bpr.2022.210036}.

\bibitem{hall1989functional}
{\sc A.~J. Hall and G.~Wake}, {\em A functional differential equation arising
  in modelling of cell growth}, J Asut Math Soc, 30 (1989), pp.~424--435,
  \url{https://doi.org/10.1017/s0334270000006366}.

\bibitem{Iyer-Biswas:2014aa}
{\sc S.~Iyer-Biswas, C.~S. Wright, J.~T. Henry, K.~Lo, S.~Burov, Y.~Lin, G.~E.
  Crooks, S.~Crosson, A.~R. Dinner, and N.~F. Scherer}, {\em {Scaling laws
  governing stochastic growth and division of single bacterial cells}}, Proc
  Natl Acad Sci USA, 111 (2014), pp.~15912--15917,
  \url{https://doi.org/10.1073/pnas.1403232111}.

\bibitem{James1971A}
{\sc James, W., Sinko, William, and Streifer}, {\em A model for population
  reproducing by fission}, Ecology, 52 (1971), pp.~330--335,
  \url{https://doi.org/10.2307/1934592}.

\bibitem{jia2021cell}
{\sc C.~Jia, A.~Singh, and R.~Grima}, {\em {Cell size distribution of lineage
  data: analytic results and parameter inference}}, iScience, 24 (2021),
  p.~102220, \url{https://doi.org/10.1016/j.isci.2021.102220}.

\bibitem{jia2022Characterizing}
{\sc C.~Jia, A.~Singh, and R.~Grima}, {\em Characterizing non-exponential
  growth and bimodal cell size distributions in fission yeast: An analytical
  approach}, PLoS Comput Biol, 18 (2022), p.~e1009793,
  \url{https://doi.org/10.1371/journal.pcbi.1009793}.

\bibitem{jones2023characterization}
{\sc I.~Jones, L.~Dent, T.~Higo, T.~Roumeliotis, M.~Arias~Garcia, H.~Shree,
  J.~Choudhary, M.~Pedersen, and C.~Bakal}, {\em Characterization of
  proteome-size scaling by integrative omics reveals mechanisms of
  proliferation control in cancer}, Sci Adv, 9 (2023), p.~eadd0636,
  \url{https://doi.org/10.1126/sciadv.add0636}.

\bibitem{jun2015cell}
{\sc S.~Jun and S.~Taheri-Araghi}, {\em Cell-size maintenance: universal
  strategy revealed}, Trends Microbiol, 23 (2015), pp.~4--6,
  \url{https://doi.org/10.1016/j.tim.2014.12.001}.

\bibitem{koch1962model}
{\sc A.~Koch and M.~Schaechter}, {\em A model for statistics of the cell
  division process}, J Gen Microbiol, 29 (1962), pp.~435--454,
  \url{https://doi.org/https://doi.org/10.1099/00221287-29-3-435}.

\bibitem{Kwon:2001aa}
{\sc C.~H. Kwon, X.~Zhu, J.~Zhang, L.~L. Knoop, R.~Tharp, R.~J. Smeyne, C.~G.
  Eberhart, P.~C. Burger, and S.~J. Baker}, {\em {Pten regulates neuronal soma
  size: a mouse model of Lhermitte-Duclos disease.}}, Nat Genet, 29 (2001),
  pp.~404--411, \url{https://doi.org/10.1038/ng781}.

\bibitem{laplante2012mTOR}
{\sc M.~Laplante and D.~Sabatini}, {\em {mTOR signaling in growth control and
  disease}}, Cell, 149 (2012), pp.~274--293,
  \url{https://doi.org/10.1016/j.cell.2012.03.017}.

\bibitem{Latronico:2004aa}
{\sc M.~V.~G. Latronico, S.~Costinean, M.~L. Lavitrano, C.~Peschle, and
  G.~Condorelli}, {\em {Regulation of cell size and contractile function by AKT
  in cardiomyocytes.}}, Ann N Y Acad Sci, 1015 (2004), pp.~250--260,
  \url{https://doi.org/10.1196/annals.1302.021}.

\bibitem{Lloyd:2013aa}
{\sc A.~C. Lloyd}, {\em {The regulation of cell size.}}, Cell, 154 (2013),
  pp.~1194--1205, \url{https://doi.org/10.1016/j.cell.2013.08.053}.

\bibitem{locasale2011Metabolic}
{\sc W.~Locasale, Jason and C.~Cantley, Lewis}, {\em Metabolic flux and the
  regulation of mammalian cell growth.}, Cell Metab, 14 (2011), pp.~443--451,
  \url{https://doi.org/10.1016/j.cmet.2011.07.014}.

\bibitem{meizlish2021tissue}
{\sc M.~L. Meizlish, R.~A. Franklin, X.~Zhou, and R.~Medzhitov}, {\em {Tissue
  Homeostasis and Inflammation}}, Annu Rev Immunol, 39 (2021), pp.~557--581,
  \url{https://doi.org/10.1146/annurev-immunol-061020-053734}.

\bibitem{Miotto2024A}
{\sc M.~Miotto, S.~Scalise, M.~Leonetti, G.~Ruocco, G.~Peruzzi, and G.~Gosti},
  {\em {A size-dependent division strategy accounts for leukemia cell size
  heterogeneity.}}, Commun Phys, 7 (2024), p.~248,
  \url{https://doi.org/10.1038/s42005-024-01743-1}.

\bibitem{m1925applications}
{\sc A.~M'kendrick}, {\em {Applications of Mathematics to Medical Problems}}, P
  Edinburgh Math Soc, 44 (1925), pp.~98--130,
  \url{https://doi.org/10.1017/S0013091500034428}.

\bibitem{Modi:2017aa}
{\sc S.~Modi, C.~A. Vargas-Garcia, K.~R. Ghusinga, and A.~Singh}, {\em
  {Analysis of noise mechanisms in cell-size control.}}, Biophys J, 112 (2017),
  pp.~2408--2418, \url{https://doi.org/10.1016/j.bpj.2017.04.050}.

\bibitem{Monds:2014aa}
{\sc R.~D. Monds, T.~K. Lee, A.~Colavin, T.~Ursell, S.~Quan, T.~F. Cooper, and
  K.~C. Huang}, {\em {Systematic perturbation of cytoskeletal function reveals
  a linear scaling relationship between cell geometry and fitness.}}, Cell Rep,
  9 (2014), pp.~1528--1537, \url{https://doi.org/10.1016/j.celrep.2014.10.040}.

\bibitem{nieto2021continuous}
{\sc C.~Nieto, C.~Vargas-Garcia, and J.~M. Pedraza}, {\em Continuous rate
  modeling of bacterial stochastic size dynamics}, Phys Rev E, 104 (2021),
  p.~044415, \url{https://doi.org/10.1103/PhysRevE.104.044415}.

\bibitem{POWELL1964A}
{\sc P.~E. O.}, {\em {A Note on Koch and Schaechter's Hypothesis about Growth
  and Fission of Bacteria}}, J Gen Microbiol, 37 (1964), p.~231,
  \url{https://doi.org/10.1099/00221287-37-2-231}.

\bibitem{otto2010signalling}
{\sc A.~Otto and K.~Patel}, {\em Signalling and the control of skeletal muscle
  size}, Exp Cell Res, 316 (2010), pp.~3059--3066,
  \url{https://doi.org/10.1016/j.yexcr.2010.04.009}.

\bibitem{palumbo2016whi5}
{\sc P.~Palumbo, M.~Vanoni, V.~Cusimano, S.~Busti, F.~Marano, C.~Manes, and
  L.~Alberghina}, {\em {Whi5 phosphorylation embedded in the G1/S network
  dynamically controls critical cell size and cell fate}}, Nat Commun, 7
  (2016), p.~11372, \url{https://doi.org/10.1038/ncomms11372}.

\bibitem{Philipp2018Analysis}
{\sc T.~Philipp}, {\em {Analysis of cell size homeostasis at the single-cell
  and population level}}, Front Phys, 6 (2018), p.~64,
  \url{https://doi.org/10.3389/fphy.2018.00064}.

\bibitem{Proulx-Giraldeau:2022aa}
{\sc F.~Proulx-Giraldeau, J.~M. Skotheim, and P.~Fran{\c c}ois}, {\em
  {Evolution of cell size control is canalized towards adders or sizers by cell
  cycle structure and selective pressures.}}, eLife, 11 (2022),
  \url{https://doi.org/10.7554/eLife.79919}.

\bibitem{Rhind:2021aa}
{\sc N.~Rhind}, {\em Cell-size control.}, Curr Biol, 31 (2021),
  pp.~R1414--R1420, \url{https://doi.org/10.1016/j.cub.2021.09.017}.

\bibitem{schaechter1958dependency}
{\sc M.~Schaechter, O.~Maaloe, and N.~O. Kjeldgaard}, {\em {Dependency on
  medium and temperature of cell size and chemical composition during balanced
  growth of Salmonella typhimurium}}, J Gen Microbiol, 19 (1958), pp.~592--606,
  \url{https://doi.org/10.1099/00221287-19-3-592}.

\bibitem{serbanescu2020nutrient}
{\sc D.~Serbanescu, N.~Ojkic, and S.~Banerjee}, {\em {Nutrient-dependent
  trade-offs between ribosomes and division protein synthesis control bacterial
  cell size and growth}}, Cell Rep, 32 (2020), p.~108183,
  \url{https://doi.org/10.1016/j.celrep.2020.108183}.

\bibitem{soifer2016single}
{\sc I.~Soifer, L.~Robert, and A.~Amir}, {\em {Single-Cell Analysis of Growth
  in Budding Yeast and Bacteria Reveals a Common Size Regulation Strategy}},
  Curr Biol, 26 (2016), pp.~356--361,
  \url{https://doi.org/10.1016/j.cub.2015.11.067}.

\bibitem{taheri2015cell}
{\sc S.~Taheri-Araghi, S.~Bradde, J.~T. Sauls, N.~S. Hill, P.~A. Levin,
  J.~Paulsson, M.~Vergassola, and S.~Jun}, {\em {Cell-size control and
  homeostasis in bacteria.}}, Curr Biol, 25 (2015), pp.~385--391,
  \url{https://doi.org/10.1016/j.cub.2014.12.009}.

\bibitem{talia2007effects}
{\sc S.~D. Talia, J.~M. Skotheim, J.~M. Bean, E.~D. Siggia, and F.~R. Cross},
  {\em The effects of molecular noise and size control on variability in the
  budding yeast cell cycle}, Nature, 448 (2007), pp.~947--951,
  \url{https://doi.org/10.1038/nature06072}.

\bibitem{tanouchi2015noity}
{\sc Y.~Tanouchi, A.~Pai, H.~Park, S.~Huang, R.~Stamatov, N.~E. Buchler, and
  L.~You}, {\em A noisy linear map underlies oscillations in cell size and gene
  expression in bacteria}, Nature, 523 (2015), pp.~357--360,
  \url{https://doi.org/10.1038/nature14562}.

\bibitem{tchouanti2024well}
{\sc J.~Tchouanti}, {\em Well posedness and stochastic derivation of a
  diffusion-growth-fragmentation equation in a chemostat}, Stoch Partial
  Differ, 12 (2024), pp.~466--524,
  \url{https://doi.org/10.1007/s40072-023-00288-8}.

\bibitem{Teimouri:2020aa}
{\sc H.~Teimouri, R.~Mukherjee, and A.~B. Kolomeisky}, {\em {Stochastic
  mechanisms of cell-size regulation in bacteria.}}, J Phys Chem Lett, 11
  (2020), pp.~8777--8782, \url{https://doi.org/10.1021/acs.jpclett.0c02627}.

\bibitem{Turner:2012aa}
{\sc J.~J. Turner, J.~C. Ewald, and J.~M. Skotheim}, {\em {Cell size control in
  yeast.}}, Curr Biol, 22 (2012), pp.~R350--9,
  \url{https://doi.org/10.1016/j.cub.2012.02.041}.

\bibitem{Tzur2009Cell}
{\sc A.~Tzur, R.~Kafri, V.~S. Lebleu, G.~Lahav, and M.~W. Kirschner}, {\em
  {Cell growth and size homeostasis in proliferating animal cells.}}, Science,
  325 (2009), pp.~167--171, \url{https://doi.org/10.1126/science.1174294}.

\bibitem{van2018cell}
{\sc B.~van Brunt, A.~Almalki, T.~Lynch, and A.~Zaidi}, {\em On a cell division
  equation with a linear growth rate}, ANZIAM J, 59 (2018), pp.~293--312,
  \url{https://doi.org/10.1017/S1446181117000591}.

\bibitem{vargas2018cell}
{\sc C.~A. Vargas-Garcia, K.~R. Ghusinga, and A.~Singh}, {\em Cell size control
  and gene expression homeostasis in single-cells}, Curr Opin in Syst Biol, 8
  (2018), pp.~109--116, \url{https://doi.org/10.1016/j.coisb.2018.01.002}.

\bibitem{wake2000functional}
{\sc G.~Wake, S.~Cooper, H.~Kim, and B.~Van-Brunt}, {\em Functional
  differential equations for cell-growth models with dispersion},
  Communications in Applied Analysis, 4 (2000), pp.~561--574.

\bibitem{Westfall:2017gd}
{\sc C.~S. Westfall and P.~A. Levin}, {\em {Bacterial cell size: Multifactorial
  and multifaceted}}, Annu Rev Microbiol, 71 (2017), pp.~499--517,
  \url{https://doi.org/10.1146/annurev-micro-090816-093803}.

\bibitem{Willis:2017aa}
{\sc L.~Willis and K.~C. Huang}, {\em {Sizing up the bacterial cell cycle}},
  Nat Rev Microbiol, 15 (2017), pp.~606--620,
  \url{https://doi.org/10.1038/nrmicro.2017.79}.

\bibitem{wood2013pom1}
{\sc E.~Wood and P.~Nurse}, {\em Pom1 and cell size homeostasis in fission
  yeast}, Cell Cycle, 12 (2013), pp.~3417--3425,
  \url{https://doi.org/10.4161/cc.26462}.

\bibitem{xia2020PDE}
{\sc M.~Xia, C.~D. Greenman, and T.~Chou}, {\em {PDE models of adder machanisms
  in cellular proliferation}}, SIAM J Appl Math, 80 (2020), pp.~1307--1335,
  \url{https://doi.org/10.1137/19M1246754}.

\bibitem{zaidi2014model}
{\sc A.~A. Zaidi, B.~Van~Brunt, and G.~C. Wake}, {\em A model for asymmetrical
  cell division}, Math Biosci Eng, 12 (2014), pp.~491--501,
  \url{https://doi.org/10.3934/mbe.2015.12.491}.

\end{thebibliography}
\end{document}